\documentclass[aps, pra, twocolumn, superscriptaddress, groupaddress, floatfix, longbibliography]{revtex4-2}
\usepackage{graphicx}
\usepackage{bm}
\usepackage{amssymb}
\usepackage{color}
\usepackage{amsmath, mathrsfs}
\usepackage{amstext}
\usepackage{float}
\usepackage{braket}
\usepackage{cancel}
\usepackage{oubraces}
\usepackage{latexsym}
\usepackage[usenames,dvipsnames]{xcolor}
\usepackage[colorlinks=true,citecolor=Blue,linkcolor=RubineRed,urlcolor=Blue]{hyperref}
\usepackage{bbold}
\usepackage{bbm}

\usepackage{stackengine}
\usepackage{graphicx}
\allowdisplaybreaks
\DeclareFontFamily{OT1}{pzc}{}
\DeclareFontShape{OT1}{pzc}{m}{it}{<-> s * [0.900] pzcmi7t}{}
\DeclareMathAlphabet{\mathpzc}{OT1}{pzc}{m}{it}

\def\H{\hat{H}}
\def\vac{\ket{vac}}
\def\w{\omega}
\def\wk{\omega_k}
\def\s{\hat{\sigma}}
\def\sd{\hat{\sigma}^{\dagger}}
\def\k{\boldsymbol{k}}
\def\ks{\boldsymbol{k} s}
\def\ksp{\boldsymbol{k}{'} {s'}}

\def\r{\boldsymbol{r}}
\def\dip{\boldsymbol{\mu}}

\def\g{\boldsymbol{g}}
\def\ad{\hat{a}^{\dagger}}
\def\adks{\hat{a}^{\dagger}_{\boldsymbol{k} s}}
\def\adkpsp{\hat{a}^{\dagger}_{\boldsymbol{k}^{'} {s'}}}

\def\aks{\hat{a}_{\boldsymbol{k}s}}
\def\aksp{\hat{a}_{\boldsymbol{k}'{s'}}}
\def\akspp{\hat{a}_{\boldsymbol{k}''{s''}}}
\def\adkspp{\hat{a}^{\dagger}_{\boldsymbol{k}''{s''}}}
\def\gks{\boldsymbol{g}_{\boldsymbol{k}s}}
\def\gkspp{\boldsymbol{g}_{\boldsymbol{k}^{''}{s''}}}

\def\gkpsp{\boldsymbol{g}_{\boldsymbol{k}^{'}{s '}}}

\def\cH{\mathcal{H}}
\def\cF{\mathcal{F}}
\usepackage[colorinlistoftodos]{todonotes}

\begin{document}
\title{Generation of polarization-entangled photon pairs from two interacting quantum emitters}

\author{Adri\'an Juan-Delgado}
\email[]{adrianjuand1996@gmail.com}
\affiliation{Centro de Física de Materiales (CMF-MPC), CSIC-UPV/EHU, 20018, Donostia-San Sebasti\'an, Spain}
\affiliation{Department of Electricity and Electronics, University of the Basque Country (UPV/EHU), Leioa 48940, Spain}

\author{Geza Giedke}
\affiliation{Donostia International Physics Center (DIPC), 20018, Donostia-San Sebasti\'an, Spain}
\affiliation{Ikerbasque, Basque Foundation for Science, 48009 Bilbao, Spain.}

\author{Javier Aizpurua}
\affiliation{Department of Electricity and Electronics, University of the Basque Country (UPV/EHU), Leioa 48940, Spain}
\affiliation{Donostia International Physics Center (DIPC), 20018, Donostia-San Sebasti\'an, Spain}
\affiliation{Ikerbasque, Basque Foundation for Science, 48009 Bilbao, Spain.}

\author{Ruben Esteban}
\email[]{ruben.esteban@ehu.eus}
\affiliation{Centro de Física de Materiales (CMF-MPC), CSIC-UPV/EHU, 20018, Donostia-San Sebasti\'an, Spain}
\affiliation{Donostia International Physics Center (DIPC), 20018, Donostia-San Sebasti\'an, Spain}

\date{\today}

\begin{abstract}
Entangled photon pairs are key elements in quantum communication and quantum cryptography. State-of-the-art sources of entangled photons are mainly based on parametric down-conversion from
nonlinear crystals, which is probabilistic in nature, and on cascade emission from biexciton quantum dots, which finds difficulties in generating entangled photons in the visible regime. Here, we use the Wigner-Weisskopf theory to provide a demonstration that polarization-entangled photon pairs can be emitted from two interacting quantum emitters with two-level-system behavior and perpendicular transition dipole moments. These emitters can represent a large variety of systems (e.g., organic molecules, quantum dots, and diamond color centers) offering a large technological versatility, for example, in the spectral regime of the emission. We show that a highly entangled photon pair can be postselected from this system by including optical filters. Additionally, we verify that the photon entanglement is not significantly affected by small changes in the detection directions and in the orientation between the dipole moments.
\end{abstract} 


\maketitle

\section{Introduction} \label{Section:Introduction}
Nonlocal quantum correlations between different systems are one of the fundamental resources in quantum technologies. These correlations are commonly referred to as \textit{entanglement} and find applications in quantum communication \cite{Bennett_PRL_1993, Bouwmeester_Nature_1997, Kimble_Nature_2008}, quantum cryptography \cite{Ekert_PRL_1991, Scarani_RMP_2009,Pironio_Nature_2010, Nadlinger_Nature_2022, Zhang_Nature_2022}, and quantum sensing and imaging \cite{Ono_NAtureCOmms_2013,Bennett_SciAdv_2016, Muller_PRL_2017, Moreau_NatureRP_2019, Toninelli_Optica_2019, Defienne_SciAd_2019, Camphausen_SciAd_2021}, among others. Photons are promising candidates for processing and distributing entanglement because they can travel long distances without being significantly affected by decoherence. For example, polarization-entangled photons generated from the cascade emission from calcium and mercury atomic beams were used in pioneering experiments testing the violation of Bell inequalities \cite{Kocher_PRL_1967, Freedman_PRL_1972, Clauser_PRL_1976, Fry_PRL_1976, Aspect_PRL_1981, Aspect_PRL_1982}. However, the radiative emission from these atomic beams is isotropic due to the randomness of the orientations of the transition dipole moments of the atoms, which reduces their technological utility. Strong efforts have thus been spent in the last few decades to design practical sources of entangled photons \cite{Edamatsu_JJA_2007, Orieux_RRP_2017}.

Nowadays, the most popular sources of entangled-photon pairs are based on parametric-down conversion (PDC) and quantum dots (QDs). On the one hand, PDC is a nonlinear optics process where a photon pumps a nonlinear crystal giving rise to the scattering of two photons. The scattered photons can be detected entangled in polarization by using a postselection procedure \cite{Shih_PRL_1988, Ou_PRL_1988, Kiess_PRL_1993} or by placing the detectors at particular directions \cite{Kwiat_PRL_1995, Kwiat_PRA_1999}. Even if the walkoff of the entangled photons generated from PDC is avoided \cite{Fedrizzi_OE_2007,Evans_PRL_2010}, they suffer from several drawbacks, such as their probabilistic nature and their large spectral linewidth. On the other hand, the cascade emission from a biexciton QD can also give rise to the emission of polarization-entangled photons \cite{Benson_PRL_2000, Stevenson_Shields_2006, Akopian_PRL_2006}, in a similar way to the cascade emission from atomic beams but with a better control on the directions of emission. 
However, the generation of entangled photons from biexciton QDs also suffers from drawbacks, such as being typically limited to the infrared range and the usual fine structure splitting that can reduce the photon entanglement \cite{Gammon_PRL_1996,Huber_PRL_2018,Liu_NatureNano_2019}. 

Notably, a source of entangled photons in the visible range can find applications in different contexts. In quantum communication, such a source would facilitate the interfacing between light and quantum nodes with optical transition frequencies \cite{Inlek_PRL2017,Drmota_PRL_2023}, and it could also facilitate the quantum-enhanced imaging of biological samples \cite{Taylor_PR_2016, Camphausen_OpticsExpress_2023}. However, only a few sources of entangled photons operating in the visible regime have been proposed \cite{Trebbia_PRA_2010, Rezai_Optica_2019, Wang_PRR_2020}, aside from the probabilistic PDC \cite{Shih_PRL_1988, Sciarrino_EPJS_2011, Sansa_APL_2022}.

Here, we present a source of entangled-photon pairs based on light emission from two interacting quantum emitters with two-level-system behavior. These emitters can represent a variety of systems, for instance, organic molecules, quantum dots, trapped ions, atoms and diamond-color centers. This variety of possible implementations offers large technological versatility, such as in choosing the spectral emission regime. For example, the application of our theoretical proposal in state-of-the-art experiments with interacting organic molecules at cryogenic temperatures would allow for the emission of photons in the visible regime  \cite{Hettich_Science_2002,Trebbia_NatComms_2022, Lange_NaturePhysics_2023, JuanDelgado_arxiv_2025}. 
We show that a highly entangled two-photon state can be postselected when the transition dipole moments of the two interacting quantum emitters are oriented perpendicularly to each other. This postselection procedure consists in including optical filters and detecting light at the direction normal to the dipole moments. Our calculations are based on the use of the Wigner-Weisskopf approximation to obtain the quantum state of the electromagnetic field, assuming that this field interacts with two initially inverted quantum emitters. 

\begin{figure}[!t] 
	\begin{center}
		\includegraphics[width=0.38\textwidth]{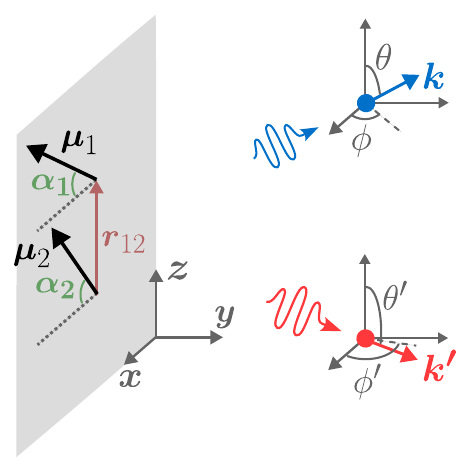}%
		\caption{Schematic representation of the two-photon emission from two initially inverted quantum emitters. The emitters (indexed by $j=1,2$) behave as two-level systems, with transition dipole moment $\dip_j = \mu ( \cos\alpha_j \hat{\boldsymbol{x}} + \sin\alpha_j \hat{\boldsymbol{z}})$, and they are located at positions $\boldsymbol{r}_j$, with $\boldsymbol{r}_{12}=\boldsymbol{r}_{1}-\boldsymbol{r}_{2}$ oriented in the $z$-direction (axis indicated at left bottom). The relaxation of the emitters generates two photons in electromagnetic modes ($\k$, $s$) and ($\k'$, $s'$) with probability amplitude $c_{\ks , \ksp}^{gg}$, where $s$ and $s'$ are the polarization modes and $\k$ and $\k'$ are the wave vectors. Additionally, $\theta$ and $\phi$ represent the polar and azimuthal angles, respectively, of the wave vector $\k$ in spherical coordinates. In Secs. \ref{Section:Entanglement_generation} and \ref{Section:postselection} we focus on the case of perpendicular transition dipole moments, with $\alpha_1 = -\alpha_2 = \pi/4$.}  
        \label{Figure:1}  
	\end{center}
\end{figure}

\section{System and model}\label{Section:Model}
In this section, we introduce the two-photon state generated from the relaxation of two quantum emitters with two-level-system behavior. These emitters are indexed by $j=1,2$ and have electronic ground state $\ket{g}_j$, excited state $\ket{e}_j$, and identical transition frequency $\omega_0$. We consider that the emitters are located at positions $\r_j$ within a homogeneous medium with refractive index $n$. Without loss of generality, we assume $\r_{12}=\r_1 - \r_2 = r_{12} \hat{\boldsymbol{z}}$, with $\hat{\boldsymbol{z}}$ the unit vector along the $z$-axis. For simplicity, we also assume that the transition dipole moments of the emitters are contained in the $xz$-plane and have identical norm $\mu$, such that $\dip_j = \mu ( \cos\alpha_j \hat{\boldsymbol{x}} + \sin\alpha_j \hat{\boldsymbol{z}})$, as schematically depicted in Fig. \ref{Figure:1}. The Hamiltonian of the quantum emitters can be written as 
\begin{equation} \label{Eq:H_QE}
    \H_{QE} = \sum_j \hbar\omega_0 \sd_j \s_j ,
\end{equation}
with $\sd_j = \ket{e}_j\bra{g}_j$ and $\s_j = \ket{g}_j \bra{e}_j$ the raising and lowering operators of emitter $j$. 

Moreover, the Hamiltonian of the electromagnetic field in the homogeneous medium can be written as an infinite summation of harmonic oscillators of frequency $\w_k$ \cite{Milonni_book}
\begin{equation} \label{Eq:EM_Hamiltonian}
    \H_{EM}= \sum_{\k, s} \hbar \w_k (\adks \aks+1/2) ,
\end{equation}
where $\adks$ and $\aks$ are the creation and annihilation operators of photons in mode ($\k$,$s$). Here, 
\begin{equation} \label{Eq:wave vector}
    \k = k (\sin\theta \cos\phi \hat{\boldsymbol{x}} +\sin\theta \sin\phi \hat{\boldsymbol{y}} + \cos\theta \hat{\boldsymbol{z}})
\end{equation}
is the photon wave vector in the homogeneous medium, where $\theta$ and $\phi$ are, respectively, the polar and azimuthal angles (see Fig. \ref{Figure:1}) and $k=\omega_k /c$, with $c$ the speed of light in the host medium with refractive index $n$. Additionally, $s$ specifies one of the two orthogonal polarization modes of the wave vector $\k$, with corresponding unit vector denoted by $\hat{\boldsymbol{e}}_{\ks}$. Furthermore, we consider that the interaction between the quantum emitters and the electromagnetic field is described by the multipolar interaction Hamiltonian \cite{Cohen_book_photons,Mandel_book_1995,Agarwal_book_2006, Steck_book_2007}
\begin{equation} \label{Eq:interaction_Hamiltonian}
\H_{I} = \hbar\sum_{\k, s} \sum_j \dip_j \cdot \gks^{(j)} \sd_j \aks + \text{H.c. },
\end{equation}
 where $\text{H.c.}$ denotes the Hermitian conjugate and the rotating-wave approximation (RWA) is used (see Appendix~\ref{Appendix:WWA} for a further discussion on the RWA). Here, we have introduced the coupling coefficient 
\begin{equation} \label{Eq:mode_function}
    \gks^{(j)} = -i \sqrt{\frac{\omega_k}{2 \varepsilon_0 n^2 \hbar \mathcal{V}}} \hat{\boldsymbol{e}}_{\k s} e^{i \k \cdot \r_j} ,
\end{equation}
with $\mathcal{V}$ the normalization volume of the electromagnetic field and $\varepsilon_0$ the vacuum permittivity.

The Wigner-Weisskopf approximation (WWA) allows us to solve the dynamics of the quantum state of the total system, starting from the initial state $\ket{\psi(0)}=\ket{e e}\vac$, in which the two emitters are in the excited state (i.e., initially inverted) and the electromagnetic field is in the vacuum state $\vac$ (i.e., no photons are present in the field). Here, we have introduced the notation $\ket{e e}=\ket{e}_1 \ket{e}_2$, where the first and second indices in $\ket{e e}$ label the state of emitter $j=1$ and $2$, respectively. In contrast with previous works that assumed parallel transition dipole moments \cite{Ernst_PR_1968,Raymond_LP_2007} or a single initially inverted emitter \cite{Milonni_PRA_1974,Svidzinsky_PRA_2010}, we consider that the two emitters are initially inverted and their transition dipole moments are oriented arbitrarily within the same $xz$-plane. We make the ansatz 
\begin{equation} \label{Eq:ansatz}
\begin{split}
    \ket{\psi (t)} &= c^{ee}(t) \ket{ee}\vac  \\
    &+ \sum_{\k, s} \biggr( c_{\k s}^{eg}(t) \ket{eg} + c_{\k s}^{ge} (t) \ket{ge} \biggr) \adks \vac \\
    &+ \sum_{\k, s} \sum_{\substack{(\k', s')\\ \geq(\k, s)}} c_{\k s , \k' {s'}}^{gg} (t)\ket{gg} \adks \adkpsp \vac ,
\end{split}
\end{equation}
which only contains terms with two excitations in total (between photons and emitter excitations), as the RWA is used. $c^{ee}(t)$ is the probability amplitude of finding the system at time $t$ still in the initial state $\ket{e e}\vac$, which satisfies $c^{ee}(0)=1$. $c_{\k s}^{eg}(t)$ and $c_{\k s}^{ge}(t)$ are the probability amplitudes of states in which, respectively, the second or the first emitter have relaxed, leading to the generation of a photon in mode ($\k$, $s$). Finally, $c_{\k s , \k' {s'}}^{gg} (t)$ is the probability amplitude of a state in which both emitters have relaxed, giving rise to two photons, one each in modes ($\k$, $s$) and ($\k'$, $s'$). We are mainly interested in the analytical expression of this two-photon probability amplitude $c_{\k s , \k' {s'}}^{gg} (t)$, as it contains all the information of the two-photon emission. 
We remark that in the double summation in the last line of Eq.~(\ref{Eq:ansatz}) each state appears and is counted only once \cite{Ernst_PR_1968}, which is indicated in the second summation of the last line in Eq.~(\ref{Eq:ansatz}) by the compact notation $(\k', s')\geq(\k, s)$. In other words, only one of the terms $\ket{gg}\ad_{\ks}\ad_{\ksp}\vac$ and $\ket{gg}\ad_{\ksp}\ad_{\ks} \vac$ appears in this summation because they represent the same physical state, as $[\adks,\adkpsp]=0$. 

Next, to obtain the analytical expressions of the probability amplitudes in Eq.~(\ref{Eq:ansatz}) we use the Schrödinger equation in the interaction picture. In this way, we find a set of coupled differential equations for these probability amplitudes, which then we solve considering that the dynamics of the system is Markovian (see Appendix~\ref{Appendix:WWA} for the complete derivation). At sufficiently long times ($t\rightarrow \infty$), we find that both emitters have relaxed [i.e., $c^{ee} (\infty)=c^{eg}_{\k s}(\infty)=c^{ge}_{\k s}(\infty)=0$] and the two-photon state becomes
\begin{equation}
 \ket{\psi (\infty)} = \sum_{\k,s}\sum_{\substack{(\k', s')\\ \geq(\k, s)}}c_{\k s , \k' {s'}}^{gg} (\infty) \ket{gg} \adks \adkpsp \vac  ,  
\end{equation}
with the two-photon probability amplitudes given by
\begin{widetext}
\begin{equation} \label{Eq:cgg_steady_state}
        c_{\ks, \ksp}^{gg}(\infty) = -\frac{S_{\ks}^{(-)} \biggr[ S_{\ksp}^{(+)} - S_{\ksp k}^{(0)}\biggr] +
           S_{\ksp}^{(-)} \biggr[S_{\ks}^{(+)}-S_{\ks k'}^{(0)}\biggr]}{2 \epsilon(\ks, \ksp)} +  \frac{A_{\ks}^{(+)}\biggr[ A_{\ksp}^{(-)} -A_{\ksp k}^{(0)}\biggr]+ A_{\ksp}^{(+)}\biggr[ A_{\ks}^{(-)} -A_{\ks k'}^{(0)}\biggr]}{2 \epsilon(\ks, \ksp)} .
\end{equation}
\end{widetext}
These probability amplitudes carry all the information of the two-photon emission: directions of emission, frequencies, and polarizations. As a consequence, they encode as well the information of the degree of two-photon entanglement. $\epsilon (\ks , \ksp)$ is the Einstein function \cite{Ernst_PR_1968}, which is equal to $2$ if $\k=\k'$ and $s=s'$, whereas it becomes equal to $1$ in any other case. Additionally, we have introduced
\begin{equation} \label{Eq:9}
    S_{\ks}^{(\pm)}= \frac{(\dip_1 \cdot \g_{\ks}^{(1)})^* + (\dip_2 \cdot \g_{\ks}^{(2)})^*}{(\gamma_0 \pm \gamma_{12})/2 + i(\omega_0 \pm V - \omega_k)}  ,
\end{equation}
\begin{equation} \label{Eq:10}
    A_{\ks}^{(\pm)}= \frac{(\dip_1 \cdot \g_{\ks}^{(1)})^* - (\dip_2 \cdot \g_{\ks}^{(2)})^*}{(\gamma_0 \pm \gamma_{12})/2 + i(\omega_0 \pm V - \omega_k)}  ,
\end{equation}
which are Lorentzian distributions related to the emission of a single photon of resonant frequency $\omega_k =\omega_0\pm V$ at rate $\gamma_0 \pm \gamma_{12}$. Here, 
\begin{equation}
\gamma_0 = \frac{\omega_0^3 |\mu|^2}{3\pi \varepsilon_0 n^2 \hbar c^3} 
\end{equation}
is the spontaneous emission rate of each emitter in the medium of refractive index $n$,
\begin{equation} \label{Eq:coherent_coupling_V}
\begin{split}
    V&=\frac{3 \gamma_0}{4} \biggr[ -\cos\alpha_1 \cos\alpha_2 \frac{\cos(k_0 r_{12})}{(k_0 r_{12})} \\
    &+ (\cos\alpha_1 \cos\alpha_2 -2\sin\alpha_1 \sin\alpha_2) \\
    &\times\biggr(\frac{\sin(k_0 r_{12})}{(k_0 r_{12})^2} + \frac{\cos(k_0 r_{12})}{(k_0 r_{12})^3 }\biggr) \biggr]
\end{split}
\end{equation}
is the coherent dipole-dipole coupling between the two emitters, and
\begin{equation}
\begin{split} \label{Eq:dissipative_coupling}
    \gamma_{12}&=\frac{3 \gamma_0}{2} \biggr[ \cos\alpha_1 \cos\alpha_2 \frac{\sin(k_0 r_{12})}{(k_0 r_{12})} \\
    &+ (\cos\alpha_1 \cos\alpha_2 -2\sin\alpha_1 \sin\alpha_2)\\
    &\times\biggr(\frac{\cos(k_0 r_{12})}{(k_0 r_{12})^2} - \frac{\sin(k_0 r_{12})}{(k_0 r_{12})^3 }\biggr) \biggr]
\end{split}
\end{equation}
is the dissipative coupling between the two emitters, with $k_0 = \w_0 /c$. The spontaneous emission rate $\gamma_0$, the coherent coupling $V$ and the dissipative coupling $\gamma_{12}$ are induced by the interaction of the emitters with the electromagnetic field in the homogeneous medium \cite{Ernst_PR_1968,Agarwal_book_2006,Stokes_NJP_2018}. Importantly, the induced dipole-dipole interaction modifies the eigenstates of the emitters, leading to new hybrid states $\ket{S}=(\ket{e g} + \ket{g e})/\sqrt{2}$ and $\ket{A}=(\ket{e g} - \ket{g e})/\sqrt{2}$ (in the absence of losses), which are symmetric and antisymmetric combinations of $\ket{e g}$ and $\ket{g e}$. The new eigenstates have energies $\hbar (\omega_0 \pm V)$, decay rates $\gamma_0 \pm \gamma_{12}$, and transition dipole moments $\dip_1 \pm \dip_2$ \cite{JuanDelgado}. 

Additionally, Eq.~(\ref{Eq:cgg_steady_state}) also includes the distributions
\begin{equation} \label{Eq:14}
        S_{\ks k'}^{(0)}= \frac{(\dip_1 \cdot \g_{\ks}^{(1)})^* + (\dip_2 \cdot \g_{\ks}^{(2)})^*}{\gamma_0 + i(2\omega_0 - \omega_{k}-\omega_{k'})}   , 
\end{equation}
\begin{equation} \label{Eq:15}
        A_{\ks k'}^{(0)}= \frac{(\dip_1 \cdot \g_{\ks}^{(1)})^* - (\dip_2 \cdot \g_{\ks}^{(2)})^*}{\gamma_0 + i(2\omega_0 - \omega_{k}-\omega_{k'})},
\end{equation}
which are related to the emission of two photons at frequencies $\omega_k$ and $\omega_{k'}$ satisfying energy conservation $\omega_k + \omega_{k'}= 2\omega_0$. These two-photon emission processes can be mediated by intermediate virtual states rather than real eigenstates of the system \cite{GonzalezTudela_NJP_2013}. 

Furthermore, the normalization condition $|\braket{\psi(\infty)|\psi(\infty)}|^2 = 1$ yields
\begin{equation}
\begin{split}
    1&=\sum_{\k, s}\sum_{\k', s'} |c_{\ks, \ksp}^{gg}(\infty)|^2 = \int d\Omega \int d\Omega' \\
    &\times\int_{0}^{\infty} dk \int_{0}^{\infty} dk' \sum_{s,s'}  \frac{k^2 (k')^2  |c_{\ks ,\ksp}^{gg}(\infty)|^2 \mathcal{V}^2}{(2\pi)^6} ,
\end{split}
\end{equation}
with $d\Omega$ and $d\Omega'$ differential solid angles that are integrated in the full space (see Appendix~\ref{Appendix:WWA}). Therefore, we can define 
\begin{equation} \label{Eq:probabolity_density_P}
    P(\k,s; \k',s') = \frac{k^2 (k')^2  |c_{\ks ,\ksp}^{gg} (\infty)|^2 \mathcal{V}^2}{(2\pi)^6} 
\end{equation}
as the probability density of emission of two photons in modes ($\k$,$s$) and ($\k'$,$s'$). We note that $P$ is independent of the choice of value of normalization volume $\mathcal{V}$, since $|c_{ks,k's}^{gg}(\infty)|^2$ scales inversely proportional with $\mathcal{V}^2$.

\section{Entanglement generation} \label{Section:Entanglement_generation}
\begin{figure*}[t] 
	\begin{center}
		\includegraphics[width=0.98\textwidth]{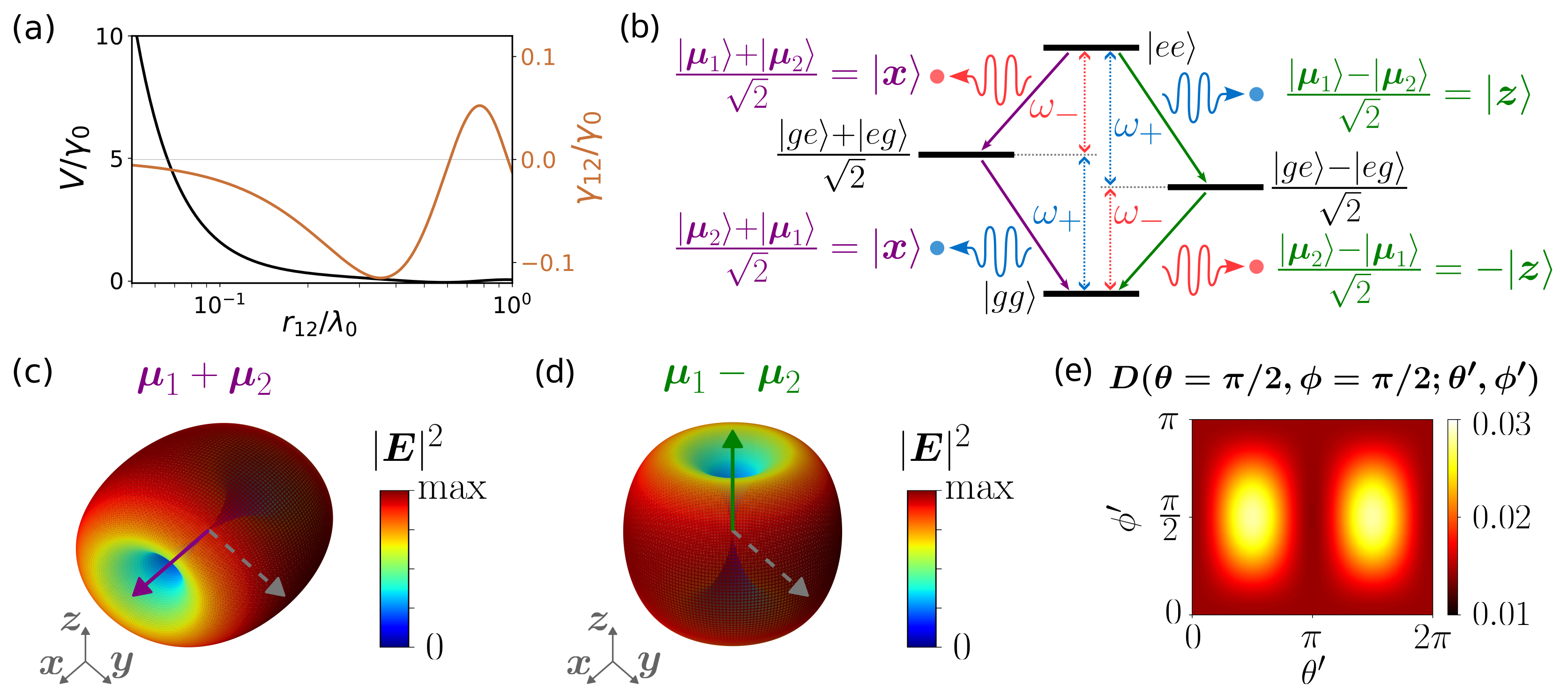}%
		\caption{Photon emission from the the symmetric and antisymmetric hybrid states. We fix $\alpha_1 = -\alpha_2 = \pi/4$ (see Fig. \ref{Figure:1}), which corresponds to perpendicular transition dipole moments. (a) Dependence on the distance $r_{12}$ between both emitters (normalized by the transition wavelength $\lambda_0= 618$ nm) of the coherent dipole-dipole coupling $V$ (black line) and of the dissipative coupling $\gamma_{12}$ (brown line). Both couplings $V$ and $\gamma_{12}$ are normalized by the spontaneous emission rate $\gamma_0$. (b) Schematic level structure and relaxation paths. The initial state $\ket{e  e}$ can relax via the symmetric state $\ket{S}=(\ket{g  e}+\ket{e g})/\sqrt{2}$ (transitions indicated with purple arrows) generating two photons polarized in the $\hat{\boldsymbol{x}}$ direction (corresponding to the direction of the transition dipole moment of the symmetric state, written in purple). $\ket{e e}$ can also relax via the antisymmetric state $\ket{A}=(\ket{g e}-\ket{e g})/\sqrt{2}$ (transitions indicated with green arrows), which leads to the emission of a photon polarized in the $\hat{\boldsymbol{z}}$ direction (opposite to the direction of the transition dipole moment of the antisymmetric state, written in green) and another photon polarized in the direction $-\hat{\boldsymbol{z}}$ (corresponding to the direction of the dipole moment of the antisymmetric state). (c) Radiation pattern of an electric point dipole oriented in the direction $\dip_1 + \dip_2 \propto \hat{\boldsymbol{x}}$ of the transition dipole moment of the symmetric state. $|\boldsymbol{E}|^2$ is the squared amplitude of the classical electric field generated by such electric-point dipole. (d) Radiation pattern of an electric point dipole oriented in the direction $\boldsymbol{\mu}_2 - \boldsymbol{\mu}_1 \propto -\hat{\boldsymbol{z}}$ (or equivalently, $\dip_1 - \dip_2 \propto  \hat{\boldsymbol{z}}$) of the transition dipole moment of the antisymmetric state. The dashed gray arrows in (c,d) mark the direction of the $y$-axis. (e) Dependence of the probability density $D(\theta=\pi/2,\phi=\pi/2;\theta',\phi')$ on $\theta'$ and $\phi'$ for two DBATT molecules separated by a distance $r_{12}=0.075\lambda_0$ and embedded in naphthalene crystal with $n=1.5$. These molecules have spontaneous emission rate $\gamma_0 /(2\pi)= 21.5$ MHz and transition wavelength $\lambda_0 = 618$ nm. }  
        \label{Figure:2}  
	\end{center}
\end{figure*}
In this section we reveal that the two-photon state $\ket{\psi(\infty)}$ can be highly entangled in polarization and frequency if the transition dipole moments $\dip_1$ and $\dip_2$ of the interacting emitters are oriented perpendicularly to each other. Additionally, we show that the probability density of two-photon emission is maximized at the normal direction to these dipole moments. To this end, we provide first an intuitive argument on the generation of entangled photons, which then we verify with the help of the analytical expressions derived in Sec. \ref{Section:Model}. 

\subsection{Entanglement of the two-photon state}
There are various ways in which a two-photon state can be considered
entangled, depending on which decomposition of the full two-photon, many-mode Hilbert space we consider. For bosons (and other indistinguishable particles), several notions of entanglement are distinguished, namely, \emph{entanglement of particles} (with different
flavors) and \emph{entanglement of modes} \cite{ESBL02,Benatti2020}. We will focus here on the latter which is more practical for typical applications of
entanglement, as, for example, in quantum communication.
In this case, the single-particle space $\cH$ of the photons is split in two or more subspaces $\cH=\cH_A\oplus\cH_B\oplus\cH_R$, corresponding to the modes of Alice, Bob,
and all remaining modes (if any), respectively. Here, we have adopted the \textit{Alice} and \textit{Bob} terminology usual in quantum information and cryptography \cite{Vedral_Plenio_PRA_1998}. The Hilbert space of the system (Fock space over $\cH$) can then be decomposed as
$\cF(\cH_A)\otimes\cF(\cH_B)\otimes\cF(\cH_R)$ and the usual notion of entanglement on composite Hilbert spaces applies.

In our case, including all the frequencies, polarizations, and propagation directions of the photons, the single-particle space is
infinite-dimensional and can host a large amount of entanglement. However, we focus here on very small (two-dimensional) subspaces $\cH_A,\cH_B$ and on the postselected state we obtain when
one photon is found in each of the two subspaces because this state is most readily used and analyzed. We show that with suitable choice of the subspaces (a pair of frequencies and
propagation directions for both polarizations) the
postselected state can be close to a maximally entangled Bell state.

\subsection{Intuitive picture of the entanglement generation} \label{Section:Simple_argument}
We consider that the directions of the transition dipole moments are fixed at $\dip_1 = \mu (\hat{\boldsymbol{x}} + \hat{\boldsymbol{z}})/\sqrt{2}$ and $\dip_2 = \mu (\hat{\boldsymbol{x}} - \hat{\boldsymbol{z}})/\sqrt{2}$ (corresponding to $\alpha_1 = -\alpha_2 = \pi /4$) and that again $\boldsymbol{r}_{12}=r_{12} \hat{\boldsymbol{z}}$. In this configuration, the emitters interact coherently if the distance $r_{12}$ between them is small in comparison to the vacuum wavelength $\lambda_0$ associated to the transition frequency $\omega_0$ of the emitters [see Eq.~(\ref{Eq:coherent_coupling_V})]. As discussed in Sec. \ref{Section:Model}, the coherent interaction leads to the formation of hybrid symmetric $\ket{S}=(\ket{g e} + \ket{e g})/\sqrt{2}$ and antisymmetric $\ket{A}=(\ket{g e} - \ket{e g})/\sqrt{2}$ states with transition dipole moments $\dip_1 + \dip_2 \propto \hat{\boldsymbol{x}}$ and $\dip_2 - \dip_1 \propto -\hat{\boldsymbol{z}}$, respectively. However, according to Eq.~(\ref{Eq:dissipative_coupling}) the dissipative coupling $\gamma_{12}$ is very small (in comparison to $\gamma_0$) for perpendicular dipoles even at very short distances. As a consequence, the decay rates of the symmetric state ($\gamma_0 + \gamma_{12} \approx \gamma_0$) and of the antisymmetric state ($\gamma_0 - \gamma_{12} \approx \gamma_0$) are very similar. To illustrate this behavior, we plot in Fig. \ref{Figure:2}a the dependence of the coherent coupling $V$ (black line) and of the dissipative coupling  $\gamma_{12}$ (brown line) on the distance $r_{12}$ in this configuration of perpendicular transition dipole moments. We consider in all the paper dibenzanthanthrene (DBATT) organic molecules as reference emitters to illustrate the results. More specifically, we fix the following molecular parameters, based on experiments in Ref. \cite{Trebbia_NatComms_2022}: (i) the decay rate is $\gamma_0 /(2\pi)= 21.5$ MHz, (ii) the transition frequency $\omega_0$ corresponds to $\lambda_0 = 618$ nm, and (iii) the host medium is naphthalene, with refractive index $n=1.5$. We neglect the influence of the combined Debye-Waller/Franck-Condon factor that effectively accounts for the effect of phonons of the host medium and of internal vibrations of the emitters \cite{JuanDelgado, Trebbia_NatComms_2022, Basche_book}. We demonstrate in Appendix~\ref{Appendix:Debye-Waller} that the value of this factor does not affect the results after an adequate scaling of the intermolecular distance $r_{12}$. 

As a consequence of such interaction (with $V\neq0$ and $\gamma_{12}\approx 0$), the initial doubly excited state $\ket{e e}$ can decay with almost the same probability to the symmetric state $\ket{S}$ and to the antisymmetric state $\ket{A}$. Importantly, the radiative decay from $\ket{e e}$ to $\ket{S}$ produces a photon of frequency $\omega_- = \omega_0 - V$ and polarization $(\ket{\boldsymbol{\mu}_1 } + \ket{\boldsymbol{\mu}_2})/\sqrt{2} = \ket{\boldsymbol{x}}$, which is followed by the relaxation from $\ket{S}$ to $\ket{g g}$ that leads to the emission of a photon of frequency $\omega_+ = \omega_0 + V$ and identical polarization $(\ket{\boldsymbol{\mu}_2 } + \ket{\boldsymbol{\mu}_1})/\sqrt{2} = \ket{\boldsymbol{x}}$. This cascade emission is schematically indicated with purple arrows in Fig. \ref{Figure:2}b. On the other hand, the radiative decay from $\ket{e e}$ to $\ket{A}$ produces a photon of frequency $\omega_+$ and polarization $(\ket{\boldsymbol{\mu}_1 } - \ket{\boldsymbol{\mu}_2})/\sqrt{2} = \ket{\boldsymbol{z}}$, which is followed by the relaxation from $\ket{A}$ to $\ket{g g}$ that leads to the emission of a photon of frequency $\omega_-$ and polarization $(\ket{\boldsymbol{\mu}_2 } - \ket{\boldsymbol{\mu}_1})/\sqrt{2} = -\ket{\boldsymbol{z}}$, see the green arrows in Fig. \ref{Figure:2}b. Therefore, this simple analysis suggests that the two-photon state is given as
\begin{equation} \label{Eq:photon_bell_state}
    \ket{\psi_-}= \frac{\ket{\hat{\boldsymbol{x}}, \omega_+}\ket{\hat{\boldsymbol{x}},\omega_-}-\ket{\hat{\boldsymbol{z}}, \omega_+}\ket{\hat{\boldsymbol{z}}, \omega_-}}{\sqrt{2}},
\end{equation}
which is entangled in frequency and polarization. 
However, we emphasize that this qualitative argument lacks information of, for example, the directions of emission. This argument also neglects the possibility of two-photon emission through intermediate virtual states, as discussed in Sec. \ref{Section:Model}.

Next, we provide a simple analysis on the directions of emission that are expected to provide a larger collection efficiency. An electric point dipole has a doughnut-shaped radiation pattern, with equal radiation strength in the plane perpendicular to the orientation of the point dipole \cite{Jackson_book_1975, Novotny_book_2012}. In Fig. \ref{Figure:2}c, we plot the radiation pattern of the transition dipole moment $\boldsymbol{\mu}_1 + \boldsymbol{\mu}_2$ of the symmetric state $\ket{S}$, which is oriented in the $x$-direction. The radiation from this dipole is maximal in the $yz$-plane. Similarly, Fig. \ref{Figure:2}d shows the radiation pattern of the transition dipole moment $\boldsymbol{\mu}_2 - \boldsymbol{\mu}_1$ of the antisymmetric state $\ket{A}$, which is oriented in the $z$-direction and has maximal radiation in the $xy$-plane. As a consequence, $\hat{\boldsymbol{y}}$ and $-\hat{\boldsymbol{y}}$ should be optimal directions of photon emission because they are directions of maximal radiation of the transition dipole moments of both hybrid states $\ket{S}$ and $\ket{A}$, see dashed gray arrows in Figs. \ref{Figure:2}c and \ref{Figure:2}d. In the following, we use the analytical expression of the two-photon probability amplitude $c_{\ks, \ksp}^{gg}(\infty)$ obtained in Sec. \ref{Section:Model} to rigorously verify these simple arguments.

\subsection{Rigorous analysis of the entanglement generation}
\begin{figure}[t] 
	\begin{center}
		\includegraphics[width=0.48\textwidth]{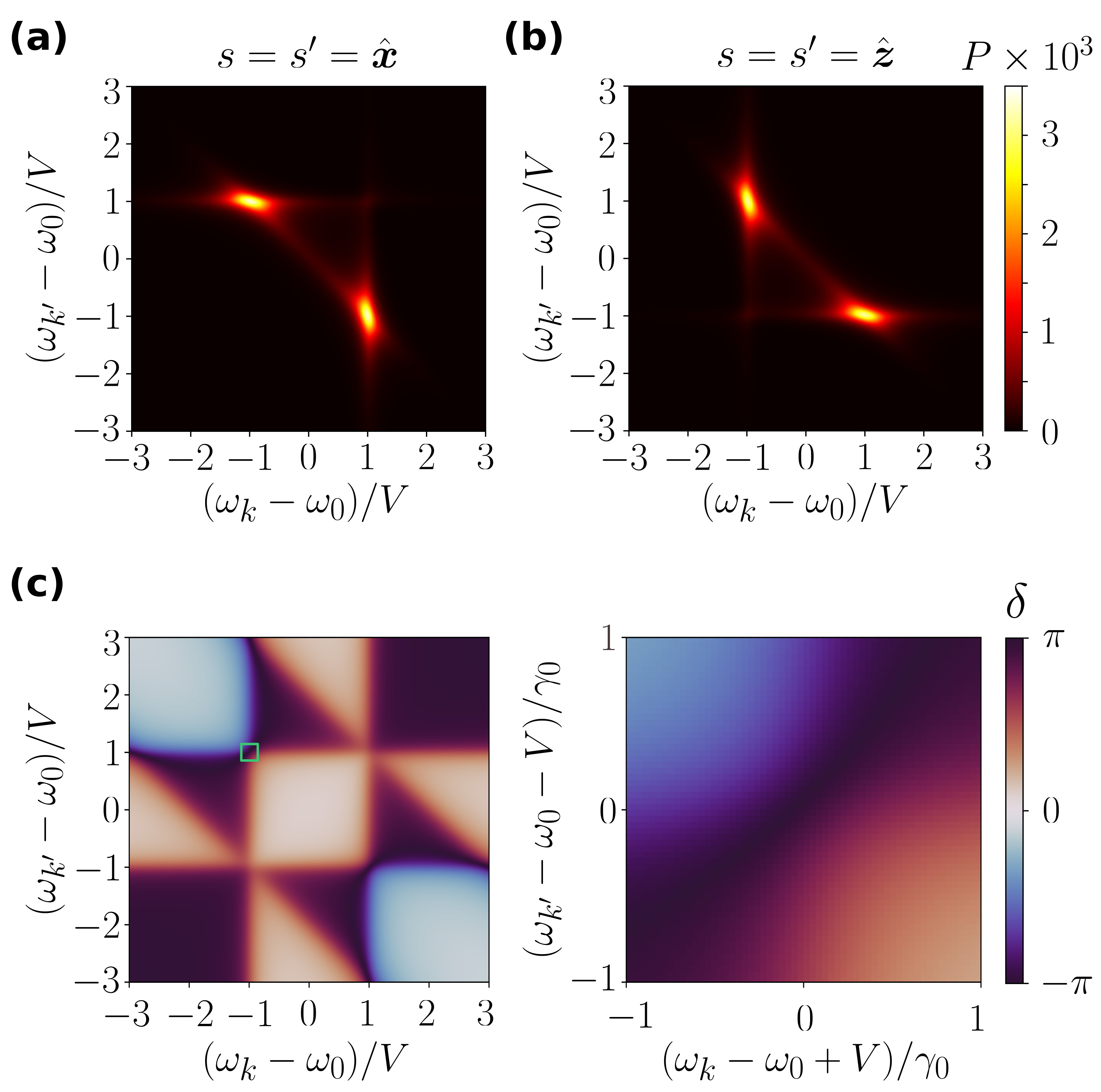}%
		\caption{Characterization of the two-photon emission at directions $\hat{\boldsymbol{k}}=\hat{\boldsymbol{y}}$ and $\hat{\boldsymbol{k}}'=-\hat{\boldsymbol{y}}$ from two DBATT molecules with perpendicular transition dipole moments. These molecules have $\gamma_0 /(2\pi)= 21.5$ MHz and $\lambda_0 = 618$ nm and they are embedded in a naphthalene crystal with refractive index $n=1.5$. Additionally, we fix the intermolecular distance at $r_{12} = 0.075 \lambda_0$ (along the z axis), which yields $V\approx 3.5\gamma_0$. We plot the probability density of two-photon emission $P(\k,s;\k',s')$ (in units of $\text{m}^2$) as a function of the photon frequencies $\omega_k =kc$ and $\omega_{k'}=k'c$ at (a) $s=s'=\hat{\boldsymbol{x}}$ and at (b) $s=s'=\hat{\boldsymbol{z}}$. (c) Relative phase $\delta$ between the two-photon probability amplitudes $c_{\ks , \ksp}^{gg}(\infty)$ at $s=s'=\hat{\boldsymbol{x}}$ and at $s=s'=\hat{\boldsymbol{z}}$. On the right panel we show a zoom of this relative phase around $\omega_k =\omega_0 - V$ and $\omega_{k'}=\omega_0 + V$ (highlighted with a green box on the left panel).}  
        \label{Figure:3}  
	\end{center}
\end{figure}
We show first that, assuming that a photon is emitted in the direction $\hat{\boldsymbol{k}}=\hat{\boldsymbol{y}}$ (normal to the $xz$-plane where the transition dipole moments are contained), the directions in which it is most likely to detect the other photon are $\hat{\boldsymbol{k}}'=\hat{\boldsymbol{y}}$ and $\hat{\boldsymbol{k}}'=-\hat{\boldsymbol{y}}$. To this end, we define $D(\theta,\phi;\theta',\phi')$ as the probability density of emission of two photons at the directions ($\theta$,$\phi$) and ($\theta'$,$\phi'$) associated with $\k$ and $\k'$, without discriminating on their frequencies and polarizations. This probability density can be calculated as
\begin{equation} \label{Eq:directivity}
    D(\theta,\phi;\theta',\phi') = \int_0^\infty dk \int_0^\infty dk' \sum_{s,s'}  P(\k,s; \k',s') .
\end{equation}
We plot in Fig. \ref{Figure:2}e this probability density as a function of $\theta'$ and $\phi'$ when we fix $\theta=\phi=\pi/2$ (corresponding to  $\hat{\boldsymbol{k}}=\hat{\boldsymbol{y}}$) and $r_{12}=0.075 \lambda_0$ (the latter yielding $V\approx 3.5 \gamma_0$). We find two regions where the probability density $D$ becomes larger, around the directions $\hat{\boldsymbol{y}}$ (at $\theta=\phi=\pi/2$) and $-\hat{\boldsymbol{y}}$ ($\theta=-\phi=\pi/2$), as expected from the simple argument in Sec. \ref{Section:Simple_argument}. We emphasize that $D(\theta,\phi;\theta',\phi')$ has been defined so that integrating this function over the four angle arguments is equal to one.

Next, we analyze the two-photon state $\ket{\psi(\infty)}$ with emission directions fixed at $\hat{\boldsymbol{k}}=\hat{\boldsymbol{y}}$ and $\hat{\boldsymbol{k}}'=-\hat{\boldsymbol{y}}$ and again fixing $r_{12}=0.075 \lambda_0$. To this aim, we plot in Fig. \ref{Figure:3}a the dependence of the probability density $P$ of photon-pair emission [Eq.~(\ref{Eq:probabolity_density_P})] on the photon frequencies $\omega_k =kc$ and $\omega_{k'}=k'c$ considering that both photons are polarized in modes $s=s'=\hat{\boldsymbol{x}}$. We find that $P$ takes maximal values $\approx 3.5 \times 10^{-3}$ $\text{m}^2$ when one of the photons has frequency $\omega_+ = \omega_0 + V$ and the other photon $\omega_- = \omega_0 - V$, corresponding to the two-photon emission via the symmetric state $\ket{S}$, see purple arrows in Fig. \ref{Figure:2}b. We find a similar dependence on $\omega_k $ and $\omega_{k'}$ of $P$ for the case in which both photons are polarized in modes $s=s'=\hat{\boldsymbol{z}}$, which is shown in Fig. \ref{Figure:3}b. The maxima have again a value of $\approx 3.5 \times 10^{-3}$ $\text{m}^2$ and are found for a photon of frequency $\omega_+$ and another photon of frequency $\omega_-$, corresponding in this case to the cascade emission via the antisymmetric state $\ket{A}$, which is indicated with green arrows in Fig. \ref{Figure:2}b. We show in Appendix~\ref{Appendix:P_xz} that $P$ drastically decreases if one of the photons has $x$-polarization and the other photon $z$-polarization, with maximum values $\approx 10^{-36}$ $\text{m}^2$. These probability densities are consistent with the entangled photon state in Eq.~(\ref{Eq:photon_bell_state}) expected from the simple analysis in Sec. \ref{Section:Simple_argument}. 

To further characterize the two-photon state $\ket{\psi(\infty)}$ at $\hat{\boldsymbol{k}}=\hat{\boldsymbol{y}}$ and $\hat{\boldsymbol{k}}'=-\hat{\boldsymbol{y}}$, we analyze the behavior of the complex phase of $c_{\ks, \ksp}^{gg}(\infty)$ at these directions. In Fig. \ref{Figure:3}c, we plot the relative phase $\delta$ between the two-photon probability amplitude at $s=s'=\hat{\boldsymbol{x}}$ and the two-photon probability amplitude at $s=s'=\hat{\boldsymbol{z}}$. More specifically,
\begin{equation} \label{Eq:relative_phase_delta}
\begin{split}
    \delta &= \text{phase}\biggr[c_{\ks, \ksp}^{gg}(\infty)\biggr]_{{\hat{\boldsymbol{k}}}=-{\hat{\boldsymbol{k}}}'=\hat{\boldsymbol{y}}, s=s'=\hat{\boldsymbol{x}}} \\
    &-\text{phase}\biggr[c_{\ks, \ksp}^{gg}(\infty)\biggr]_{{\hat{\boldsymbol{k}}}=-{\hat{\boldsymbol{k}}}'=\hat{\boldsymbol{y}}, s=s'=\hat{\boldsymbol{z}}} .
\end{split}
\end{equation}
We find $\delta\approx \pi$ near the photon frequencies that yield maximal probability density $P$ of two-photon emission (i.e., one photon at $\omega_+$ and another photon at $\omega_-$), as can be appreciated more easily in the zoom in the right panel of Fig. \ref{Figure:3}c. This difference of phase agrees with the relative phase between $\ket{\hat{\boldsymbol{x}}, \omega_-}\ket{\hat{\boldsymbol{x}},\omega_+}$ and $\ket{\hat{\boldsymbol{z}}, \omega_-}\ket{\hat{\boldsymbol{z}}, \omega_+}$ in Eq.~(\ref{Eq:photon_bell_state}). However, deviations in the photon frequencies of $\approx\gamma_0$ are sufficient to strongly modify the relative phase (which implies deviating from the maximally entangled state) and, consequently, filters with very narrow linewidths are needed to postselect a highly entangled state in the next section. 

\section{Postselection of a highly polarization-entangled state} \label{Section:postselection}
\begin{figure}[t] 
	\begin{center}
		\includegraphics[width=0.48\textwidth]{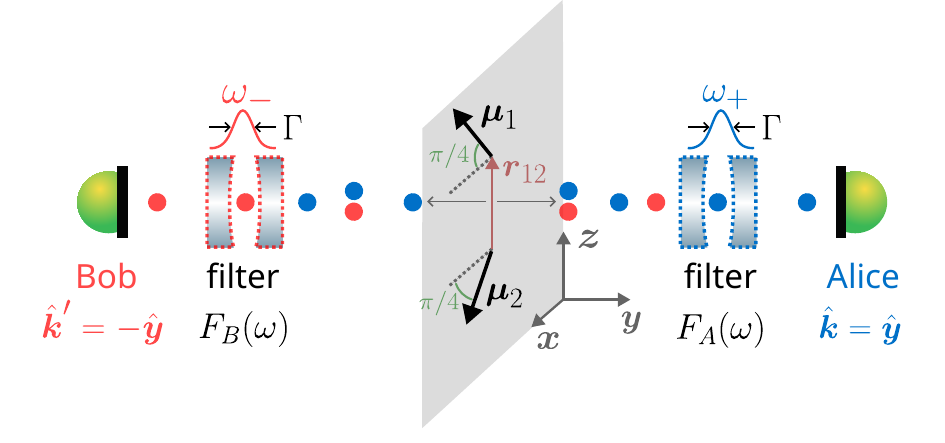}%
		\caption{Schematic representation of the postselection procedure. The transition dipole moments of the emitters are assumed to be contained in the $xz$-plane and oriented perpendicularly to each other ($\alpha_1 = -\alpha_2 = \pi/4$ in Fig. \ref{Figure:1}). Blue circles represent photons emitted at frequency $\approx\omega_+ = \omega_0 + V$ from the interacting system, while red circles correspond to photons emitted at frequency $\approx\omega_- = \omega_0 - V$. \textit{Alice} detects only photons emitted at $\hat{\boldsymbol{k}}=\hat{\boldsymbol{y}}$ and \textit{Bob} does it at $\hat{\boldsymbol{k}}'=-\hat{\boldsymbol{y}}$. Additionally, Alice uses a filter with Lorentzian profile $F_A (\omega)$, with linewidth $\Gamma$ and central frequency $\omega_+$, whereas Bob uses a filter with Lorentzian profile $F_B (\omega)$, with linewidth $\Gamma$ and central frequency $\omega_- $.}  
        \label{Figure:4}  
	\end{center}
\end{figure}
\begin{figure*}[t] 
	\begin{center}
		\includegraphics[width=0.98\textwidth]{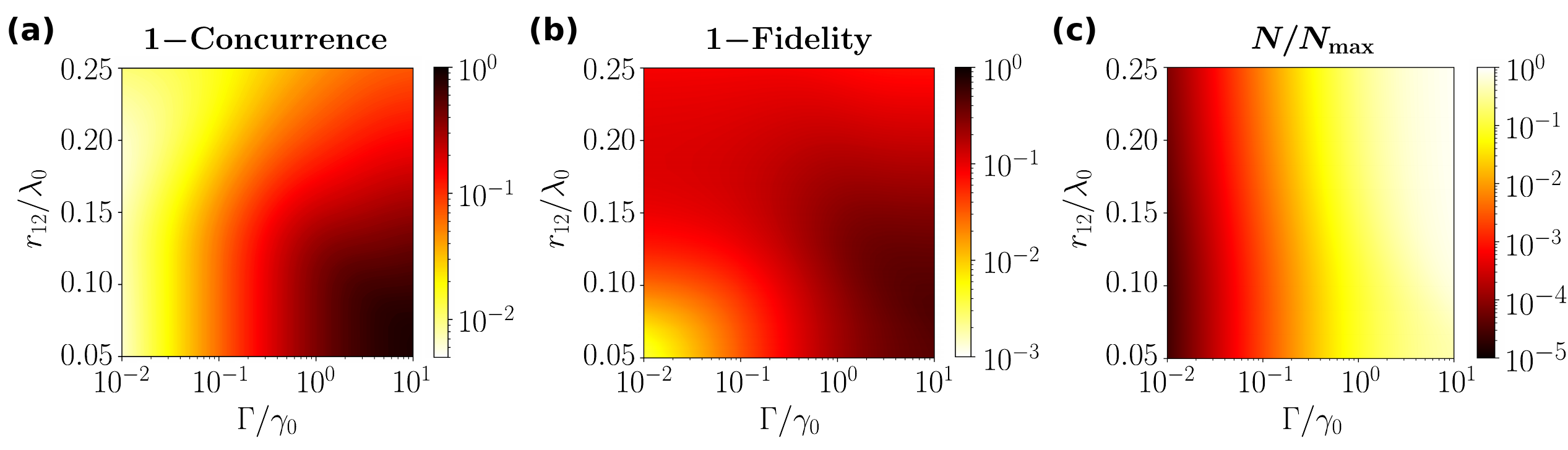}%
		\caption{Characterization of the two-photon postselected state $\hat{\rho}$. We plot the dependence on the linewidth $\Gamma$ of the filters (normalized by the spontaneous emission rate $\gamma_0$) and on the distance $r_{12}$ between the two emitters (normalized by $\lambda_0$) of (a) $1-\mathcal{C}$ (where $\mathcal{C}$ is the concurrence), (b) $1-\mathcal{F}$ (where $\mathcal{F}$ is the the fidelity of $\hat{\rho}$ with respect to the Bell state $(\ket{\hat{\boldsymbol{x}}\hat{\boldsymbol{x}}}-\ket{\hat{\boldsymbol{z}}\hat{\boldsymbol{z}}})/\sqrt{2}$), and (c) the normalizing factor $N$ of the density matrix divided by its maximum value $N_{\text{max}}$ (obtained within the range of filter linewidths and intermolecular distances analyzed). The two emitters are DBATT molecules, with $\gamma_0 /(2\pi)= 21.5$ MHz and $\lambda_0 = 618$ nm, which are embedded in a naphthalene crystal with refractive index $n=1.5$. The transition dipole moments of these molecules are contained in the $xz$-plane and oriented perpendicularly to each other ($\alpha_1 = - \alpha_2 = \pi/4$ in Fig. \ref{Figure:1}).}
        \label{Figure:5}  
	\end{center}
\end{figure*}
In this section we show that a two photon-state that is highly entangled in polarization can be postselected using optical filters. We consider that Alice detects light propagating in the direction $\hat{\boldsymbol{y}}$ and Bob does it in the direction $-\hat{\boldsymbol{y}}$, as schematically represented in Fig. \ref{Figure:4}. The postselected state is based on the detection of a single photon by Alice and a single photon by Bob and, thus, determined by the two-photon probability amplitudes $c_{\ks ,\ksp}^{gg}(\infty)$ with $\hat{\boldsymbol{k}}=\hat{\boldsymbol{y}}$ and $\hat{\boldsymbol{k}}'=-\hat{\boldsymbol{y}}$. Additionally, Alice ($A$) and Bob ($B$) use optical filters with Lorentzian profiles 
\begin{subequations}
    \begin{align}
        F_A (\omega) &= \frac{\Gamma/2}{(\Gamma/2)+i(\omega- \omega_+)}, \\
        F_B (\omega) &= \frac{\Gamma/2}{(\Gamma/2)+i(\omega- \omega_-)}. 
    \end{align}
\end{subequations}
These profiles assume that both optical filters have the same linewidth $\Gamma$, but while Alice filters light around $\omega_+$, Bob does it around $\omega_-$. 

Moreover, we consider that Alice and Bob use detectors that count all photons that pass through the filters, without distinguishing their frequency. As a result, the postselected state has only polarization degrees of freedom and is properly described by a two-photon density matrix $\hat{\rho}$ (rather than by a pure state) due to the erasing of frequency information. (We analyze the dependence of the purity of $\hat{\rho}$ on $\Gamma$ and $\r_{12}$ in Appendix~\ref{Appendix:Purity}.) To obtain the postselected state we follow the usual tomography procedure \cite{Altepeter_review_2005, Nodar_arxiv_2022}. In the orthogonal basis of polarization formed by the $\hat{\boldsymbol{x}}$ and $\hat{\boldsymbol{z}}$ directions (which are also orthogonal to the detection directions $\hat{\boldsymbol{k}}=\hat{\boldsymbol{y}}$ and $\hat{\boldsymbol{k}}'=-\hat{\boldsymbol{y}}$), the elements of the density matrix $\hat{\rho}$ are given by \cite{Nodar_arxiv_2022}
\begin{equation} \label{Eq:postselected_state}
    \begin{split}
        \bra{u u'}\hat\rho\ket{s s'}&= \frac{1}{N} \int_0^{\infty} d\omega_k \int_0^{\infty} d\omega_{k'} \braket{ u (\omega_k) u'  (\omega_{k'}) |\psi (\infty)} \\
        &\times \braket{\psi (\infty)| s (\omega_k) s' (\omega_{k'})}  , 
    \end{split}
\end{equation}
with $u,u',s,s'\in\{ \hat{\boldsymbol{x}},\hat{\boldsymbol{z}}\}$. $\ket{s (\omega_k) s' (\omega_{k'})}$ is a two-photon pure state that (i) involves a photon of frequency $\omega_k$ propagating towards Alice ($\hat{\boldsymbol{k}}=\hat{\boldsymbol{y}}$) and a photon of frequency $\omega_{k'}$ propagating towards Bob ($\hat{\boldsymbol{k}}'=-\hat{\boldsymbol{y}}$), and (ii) accounts for the influence of the optical filters. More specifically, this state is given as
\begin{equation}
\begin{split}
    \ket{s (\omega_k) s' (\omega_{k'})} &= F_A (\omega_k)F_B (\omega_{k'})  \\
    &\times\ad_{\ks}\ad_{\ksp} \vac \biggr\rvert_{{\boldsymbol{k}}=\hat{\boldsymbol{y}}\omega_k /{c} , {\boldsymbol{k}}'=-\hat{\boldsymbol{y}}\omega_{k'}/{c} } . 
\end{split}
\end{equation}
Further, in Eq.~(\ref{Eq:postselected_state}) we have included the normalization factor  
\begin{equation} \label{Eq:Normalization_factor}
    N = \sum_{u,u'}\int_0^{\infty} d\omega_k \int_0^{\infty} d\omega_{k'}  |\braket{\psi (\infty)| u (\omega_k) u' (\omega_{k'})}|^2 ,
\end{equation}
which guarantees that $\text{Tr}{\hat{\rho}}=1$. 

We quantify next the degree of entanglement of the two-photon postselected state $\hat{\rho}$ and its dependence on the distance $r_{12}$ between the emitters and on the linewidth $\Gamma$ of the filters. To this end, we compute the concurrence $\mathcal{C}(\hat{\rho})$, which measures the degree of entanglement of formation of any two-qubit system \cite{Wootters_PRL_1998}. $\mathcal{C}(\hat{\rho})$ can be obtained as
\begin{equation}
    \mathcal{C} (\hat{\rho}) = \text{max}\{0, \sqrt{\lambda_1}-\sqrt{\lambda_2}-\sqrt{\lambda_3}-\sqrt{\lambda_4}\} .
\end{equation}
Here, $\lambda_i$ are the eigenvalues (in decreasing order) of $\hat{\rho} \hat{\tilde{\rho}}$, where
\begin{equation}
    \hat{\tilde{\rho}} = (\sigma_A^y \otimes \sigma_B^y ) \hat{\rho}^* (\sigma_A^y \otimes \sigma_B^y ),
\end{equation}
with $\sigma_\chi^y = -i \ket{\hat{\boldsymbol{x}}}_\chi \bra{\hat{\boldsymbol{z}}}_\chi +i \ket{\hat{\boldsymbol{z}}}_\chi\bra{\hat{\boldsymbol{x}}}_\chi$ the $y$-Pauli matrix in the Hilbert space of the polarization of the photon detected by Alice ($\chi=A$) or Bob ($\chi=B$). Concurrence is bounded between $0$ and $1$, taking the lowest value if $\hat{\rho}$ is a separable state and the highest value if it is a maximally entangled state. Thus, $1-\mathcal{C}(\hat{\rho})$ measures the deviation of the postselected state from a maximally entangled state.

To examine the behavior of concurrence, we consider again the case of two DBATT molecules, although we have verified that equivalent results are obtained for different quantum emitters provided that the same dipole–dipole coupling strength is fixed. Figure \ref{Figure:5}a shows the dependence of $1-\mathcal{C}(\hat{\rho})$ on $\Gamma$ (normalized by $\gamma_0$) and on $r_{12}$ (normalized by $\lambda_0$). We find that filters with very narrow linewidth ($\Gamma/\gamma_0 \ll 0.1$) are needed to obtain a highly entangled postselected state ($1-\mathcal{C} \lesssim 10^{-2}$). Additionally, at $\Gamma/\gamma_0 \ll 0.1$ the dependence of the concurrence on $r_{12}$ is small for the range of distances analyzed here (see Appendix~\ref{Appendix:distant_emitters} for a discussion on larger separation distances). We attribute the necessity of very narrow filters to the high sensitivity of the relative phase $\delta$ between the two-photon probability amplitude $c_{\ks, \ksp}^{gg}(\infty)$ at $s=s'=\hat{\boldsymbol{x}}$ and the two-photon probability amplitude $c_{\ks, \ksp}^{gg}(\infty)$ at $s=s'=\hat{\boldsymbol{z}}$ [both of them evaluated at $\hat{\boldsymbol{k}}=-\hat{\boldsymbol{k}}'=\hat{\boldsymbol{y}}$, see Eq. (\ref{Eq:relative_phase_delta})]. As discussed in Sec. \ref{Section:Entanglement_generation} and shown in Fig. \ref{Figure:3}c, if one of the photons has frequency $\omega_+$ and the other one $\omega_-$ we find $\delta\approx\pi$ (corresponding to a Bell state, with maximum entanglement), but small deviations in the photon frequencies drastically change this value of relative phase. 

Moreover, to further verify the simple argument in Sec. \ref{Section:Entanglement_generation}, we analyze the similarity between the postselected state $\hat{\rho}$ and the polarization Bell state
\begin{equation}
    \ket{\psi_-^{\text{Bell}}} = \frac{\ket{\hat{\boldsymbol{x}}\hat{\boldsymbol{x}}}-\ket{\hat{\boldsymbol{z}}\hat{\boldsymbol{z}}}}{\sqrt{2}} ,
\end{equation}
expected from Eq.~(\ref{Eq:photon_bell_state}) once the frequency degrees of freedom are erased. With this objective, we compute the fidelity of $\hat{\rho}$ with respect to such state, which is given as
\begin{equation} \label{Eq:fidelity}
    \mathcal{F}(\hat{\rho}) = \bra{\psi_-^{\text{Bell}}}\hat{\rho}\ket{\psi_-^{\text{Bell}}} .
\end{equation}
Figure \ref{Figure:5}b shows the dependence on $\Gamma/\gamma_0$ and on $r_{12}/\lambda_0$ of $1-\mathcal{F}(\hat{\rho})$. We find again that very narrow filters are required to minimize the deviation of the postselected state from $\ket{\psi_-^{\text{Bell}}}$. Additionally, 
for very narrow filters ($\Gamma/\gamma_0 \ll 0.1$) we obtain that decreasing the intermolecular distances $r_{12}$ further minimizes the deviation of the postselected state from $\ket{\psi_-^{\text{Bell}}}$, which is a consequence of the relative phase $\delta$ approaching $\pi$ more closely as $r_{12}$ decreases. 

These findings indicate that filters with very narrow linewidths are required to obtain a highly polarization-entangled postselected state. However, the probability of Alice and Bob receiving a single photon each one decreases with the narrowness of the filters. To quantify how much this probability is reduced in comparison to the case in which broad filters are used, we analyze here the factor $N$, given by the trace of $\hat{\rho}$ before normalization [see Eq.~(\ref{Eq:Normalization_factor})]. (We additionally provide in Appendix \ref{Appendix:brightness} a zeroth-order estimation of the brightness of the entangled-photon source.) We plot in Fig. \ref{Figure:5}c the dependence on $\Gamma/\gamma_0$ and on $r_{12}/\lambda_0$ of $N$ divided by the maximum value $N_{\text{max}}$ obtained within the range of linewidth and intermolecular distance explored in this figure. We find that, for very narrow filters ($\Gamma\approx10^{-2} \gamma_0$), $N$ can be up to five orders of magnitude smaller than for broad filters ($\Gamma\approx10 \gamma_0$). Thus, to choose the optimal spectral widths of the filters in experiments, it is necessary to consider a balance between the two-photon entanglement and the detection probability, as decreasing values of $\Gamma$ increase the concurrence of the postselected state, but at the cost of lowering $N/N_{\text{max}}$. 

In the following we analyze the entanglement of the postselected state under different detection directions and under misaligments in the relative orientation between the transition dipole moments.

\subsection{Two-photon entanglement under different detection directions} \label{Subsection:lens}
\begin{figure}[t] 
	\begin{center}
		\includegraphics[width=0.48\textwidth]{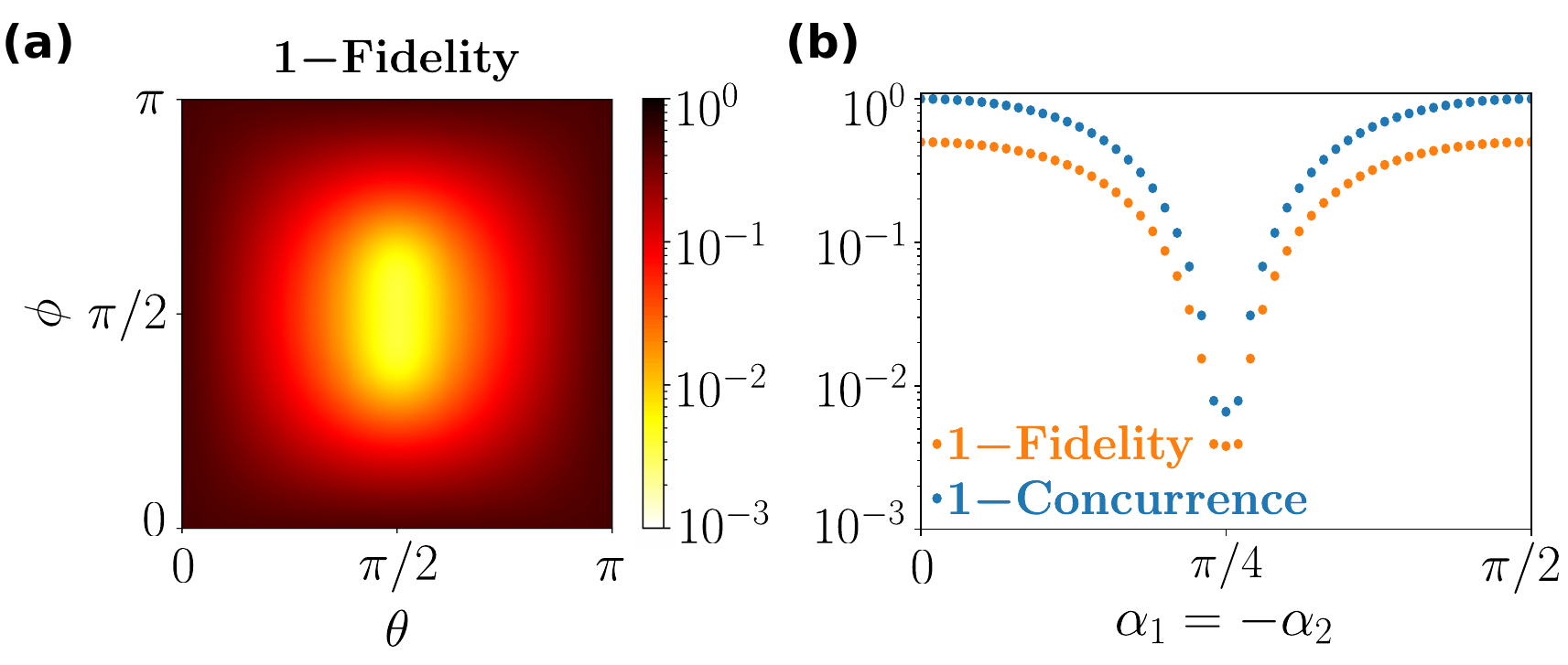}%
		\caption{Robustness of the two-photon postselected state against the detection direction and against the angle between the transition dipoles. The two emitters are DBATT molecules, with $\gamma_0 /(2\pi)= 21.5$ MHz and $\lambda_0 = 618$ nm, which are embedded inside a naphthalene crystal with refractive index $n=1.5$ and separated by a distance $r_{12}=0.05\lambda_0$. The linewidth of the filters is $\Gamma/\gamma_0 =10^{-2}$. (a) Dependence on the detection direction $\hat{\boldsymbol{k}}=\hat{\boldsymbol{k}}(\theta,\phi)$ of Alice of the deviation of the postselected state $\hat{\rho}$ from the Bell state $(\ket{\hat{\boldsymbol{x}}\hat{\boldsymbol{x}}}-\ket{\hat{\boldsymbol{z}}\hat{\boldsymbol{z}}})/\sqrt{2}$, which is quantified through $1-\mathcal{F}$. The detection direction of Bob is fixed at $\hat{\boldsymbol{k}}'=-\hat{\boldsymbol{y}}$ and the transition dipole moments are oriented perpendicularly to each other, as in previous simulations (sketch in Fig. \ref{Figure:1}, with $\alpha_1 = -\alpha_2 = \pi/4$). (b) Dependence on the angles $\alpha_1  = -\alpha_2$ of $1-\mathcal{C}$ and $1-\mathcal{F}$, at detection directions $\hat{\boldsymbol{k}}=\hat{\boldsymbol{y}}$ and $\hat{\boldsymbol{k}}'=-\hat{\boldsymbol{y}}$ normal to the $xz$-plane in which the transition dipole moments $\dip_j = \mu (\cos\alpha_j \hat{\boldsymbol{x}}+\sin\alpha_j\hat{\boldsymbol{z}})$ are contained.}  
        \label{Figure:6}  
	\end{center}
\end{figure}
In experiments, light is collected by a lens, whose numerical aperture is key (together with the filters' width) to determine the collection efficiency. A complete analysis of the influence of lenses in the photon entanglement goes beyond the scope of this work. However, to gain an understanding of the expected impact of the lens on the two-photon probability amplitudes, we consider next the effect of deviations in the detection direction. We show that moderate deviations are not expected to significantly affect the high values of fidelity $\mathcal{F}(\hat{\rho})$ of the postselected state $\hat{\rho}$ with respect to the Bell state $\ket{\psi_-^{\text{Bell}}}$. We focus on the analysis of $\mathcal{F}(\hat{\rho})$ because this quantity measures the distance of $\hat{\rho}$ from a fixed state (in our case, $\ket{\psi_-^{\text{Bell}}}$) and, thus, we expect that if the fidelity is high over all individual angles collected by a lens of a given numerical aperture, the fidelity of the state that can be measured in a straightforward way considering all the collected light will also be high. For example, if the state of the emitted light (after the lens) is $\ket{\psi_-^{\text{Bell}}}$ for all angles, we expect that the whole collected beam will be perfectly entangled and that this entanglement can be easily measured by using two polarizers.

We consider that Alice detects light propagating in some direction $\hat{\boldsymbol{k}}=\hat{\boldsymbol{k}}(\theta,\phi)$ satisfying $\hat{\boldsymbol{k}} \cdot \hat{\boldsymbol{y}}>0$, whereas the detection direction of Bob is again fixed at $-\hat{\boldsymbol{y}}$. 
Additionally, Alice and Bob measure the polarization of photons in the same basis $\{\ket{\hat{\boldsymbol{x}}},\ket{\hat{\boldsymbol{z}}}\}$. However, the polarization directions of the photons propagating towards Alice are generally different to $\hat{\boldsymbol{x}}$ and $\hat{\boldsymbol{z}}$ (except at $\theta=\phi=\pi/2$, which is the case analyzed in previous sections). To collimate the photons and rotate their directions of polarization, guaranteeing that they are polarized in the $xz$-plane, we consider that Alice uses a lens oriented normally to the $y$-axis. The effect of the lens is included through the application of a unitary transformation to the two-photon probability amplitudes of the postselected state, which is discussed in Appendix~\ref{Appendix:detection_angles}. 

To show the effect of deviation from the detection at $\hat{\boldsymbol{k}}(\theta,\phi)=\hat{\boldsymbol{y}}$, we plot in Fig. \ref{Figure:6}a the dependence of $1-\mathcal{F}(\hat{\rho})$ on $\theta$ and $\phi$ for two DBATT molecules at $r_{12}=0.05\lambda_0$ and $\Gamma=10^{-2}\gamma_0$. We find $1-\mathcal{F}(\hat{\rho})\lesssim 10^{-2}$ over a range of angles covering a large solid angle. Therefore, we expect that the integration of the emission over such solid angle would still yield a highly polarization-entangled state while increasing the collection efficiency. In Appendix \ref{Appendix:distant_emitters}, we show that, in contrast to the behavior of $\mathcal{F}(\hat{\rho})$ found here for small separation distances $r_{12}$, $\mathcal{F}(\hat{\rho})$ drastically decreases under small deviations in the detection direction when the separation $r_{12}$ is large (comparable to or larger than $\lambda_0$). This difference highlights the advantage of quantum emitters at short separation distances for practical entanglement generation.

\subsection{Robustness of the entanglement to misaligments in the orientation of the transition dipole moments} \label{Subsection:robustness}
Finally, we verify that the large values of concurrence and fidelity that we have obtained assuming transition dipole moments oriented in perpendicular directions are robust against orientation misalignments. To this end, we recall the general expressions of the transition dipole moments $\dip_j = \mu (\cos\alpha_j \hat{\boldsymbol{x}}+\sin\alpha_j\hat{\boldsymbol{z}})$. Here, we fix $\alpha_1 = -\alpha_2$ and compute the postselected state $\hat{\rho}$ for $\alpha_1 \in [0,\pi/2]$. Additionally, we consider again that Alice detects photons in the direction $\hat{\boldsymbol{k}}=\hat{\boldsymbol{y}}$ and Bob does it in the direction $\hat{\boldsymbol{k}}'=-\hat{\boldsymbol{y}}$. Figure \ref{Figure:6}b shows the dependence on $\alpha_1 =-\alpha_2$ of $1-\mathcal{C}(\hat{\rho})$ (blue dots) and $1-\mathcal{F}(\hat{\rho})$ (orange dots), which reach minimal values at $\alpha_1 = \pi/4$, corresponding to perpendicular transition dipole moments. Importantly, we find very low values of $1-\mathcal{C}(\hat{\rho})$ and $1-\mathcal{F}(\hat{\rho})$ also for moderate deviations from $\alpha_1 = \pi/4$, which indicates that the two-photon postselected state is highly entangled even if the dipoles are not exactly perpendicular.

\section{Conclusions}
In summary, we provide a demonstration that two interacting quantum emitters with two-level behavior can be used as a source of entangled photons. These quantum emitters can represent a variety of systems (for example, organic molecules, trapped ions, quantum dots, atoms, and diamond-color centers), which provides a large technological versatility. For example, emission of entangled-photon pairs in the visible range could be obtained in state-of-the-art experiments with interacting organic molecules at cryogenic temperatures \cite{Hettich_Science_2002, Trebbia_NatComms_2022, Lange_NaturePhysics_2023, JuanDelgado_arxiv_2025}.

We have derived the dynamics of the quantum state of the electromagnetic field interacting with two quantum emitters using the Wigner-Weisskopf approximation. Considering that the emitters are initially inverted and have perpendicular transition dipole moments, we have demonstrated that a highly polarization-entangled two-photon state can be postselected. More specifically, we consider that Alice and Bob are located at the normal directions to the transition dipole moments of the emitters and, additionally, each of them uses an optical filter. We have found that the entanglement (quantified through the concurrence) increases with decreasing spectral widths of the filters. Additionally, the fidelity of the postselected state with respect to a Bell state increases at very short separation distances between the emitters. Furthermore, we have verified that this fidelity is high even if light is detected with some deviation from the normal direction to the dipole moments, which indicates that lenses could provide larger collection efficiencies without disturbing significantly the photon entanglement. We have also checked that the high photon entanglement is robust against misaligments in the orientation between the transition dipole moments. 

Finally, the optimal values of spectral widths of the filters in experiments depend on the desired balance between the degree of two-photon entanglement of the postselected state (which increases with decreasing bandwidth of the filters) and the probability of two-photon detection (which decreases with the bandwidth of the filters). Future theoretical analyses could address the complete impact of lenses and finite-size detectors and/or the assistance of optical cavities to improve the collection efficiency without affecting the degree of two-photon entanglement \cite{Johne_PRL_2008,Dousse_Nature_2010}.

These results show that interacting quantum emitters can be exploited to produce highly polarization-entangled photon pairs, with large versatility and adaptability, and thus become very promising building blocks for quantum communication, cryptography, sensing and imaging.
\vfill

\section{Acknowledgments}
We thank Carlos Maciel-Escudero, \'Alvaro Nodar, Brahim Lounis, Jean-Baptiste Trebbia, Tomá\v{s} Neuman, Mikolaj K. Schmidt, Alejandro Gonzalez-Tudela and Gabriel Molina-Terriza for helpful discussions. A.J.D., R.E., and J.A. acknowledge financial support through the Grant No. PID2022-139579NB-I00 funded by MICIU/AEI/10.13039/501100011033 and by ERDF/EU, through the Grant No. IT 1526-22 funded by the Department of Science, Universities and Innovation of the Basque Government, through the Laboratory for Transborder Cooperation LTC TRANS-LIGHT from University of the Basque Country and University of Bordeaux, and also by the European Commission - NextGenerationEU (Regulation EU 2020/2094), through CSIC's Quantum Technologies Platform (QTEP) (Project No. 20219PT023). A.J.D. acknowledges financial support through the Grant No. PRE2020-095013 funded by MICIU/AEI/10.13039/501100011033 and by ``ESF Investing in your future". G.G. acknowledges funding through the Grant No. PID2023-146694NB-I00 funded by 
MICIU/AEI/10.13039/501100011033/ and by ERDF/EU, by the Basque Department of Education (Grant No. PIBA-2023-1-0021), by the European Union NextGenerationEU/PRTR-C17.I1, and financial support through the IKUR Strategy under the collaboration agreement between Ikerbasque Foundation and DIPC on behalf of the Department of Education of the Basque Government.\\

\textit{Data availability}.-- The data that support the findings of this article are openly available \cite{data_availability}.
\appendix
\section{Wigner-Weisskopf approximation} \label{Appendix:WWA}
In this appendix, we use the Wigner-Weisskopf approximation to obtain the time evolution of the two-photon probability amplitudes $c_{\ks, \ksp}^{gg}(t)$. To this end, we first derive a set of coupled differential equations for the probability amplitudes $c_{\ks, \ksp}^{gg}(t)$, $c_{\ks}^{eg}(t)$, $c_{\ks}^{ge}(t)$ and $c^{ee}(t)$ of the ansatz $\ket{\psi(t)}$ proposed in Eq.~(\ref{Eq:ansatz}) in the main text. This set of equations is obtained by substituting $\ket{\psi(t)}$ into the interaction picture Schrödinger equation
\begin{equation} \label{Eq:Schrodinger_eq}
    i\frac{d}{dt}\ket{\psi(t)} = \frac{1}{\hbar}\hat{H}_{I}(t) \ket{\psi(t)} .
\end{equation}
Here, $\hat{H}_{I}(t)$ is the interaction Hamiltonian written in the interaction picture under the rotating-wave approximation (RWA) and it is given by 
\begin{equation} \label{Eq:interaction_Hamiltonian_interaction_picture}
\begin{split}
\H_{I} (t) &= e^{i(\H_{QE} + \H_{EM})t/\hbar} \H_I e^{-i(\H_{QE} + \H_{EM})t/\hbar} \\
&= \frac{1}{\hbar}\sum_{\k, s} \sum_j \dip_j \cdot \gks^{(j)} \sd_j \aks e^{i(\omega_0 - \w_k)t} + \text{H.c.} ,
\end{split}
\end{equation}
where $\H_{QE}$, $\H_{EM}$ and $\H_I $ are the Schrödinger picture Hamiltonians of the quantum emitters, of the electromagnetic field in the homogeneous medium, and of the interaction between them, respectively [Eqs. (\ref{Eq:H_QE}), (\ref{Eq:EM_Hamiltonian}), and (\ref{Eq:interaction_Hamiltonian}) in the main text]. Additionally, $\dip_j$ is the transition dipole moment of emitter $j$, $\gks^{(j)}$ is the coupling coefficient of emitter $j$ with the electromagnetic field [Eq. (\ref{Eq:mode_function})], $\omega_0$ is the transition frequency of the emitters, $\sd_j$ is the raising operator of emitter $j$, and $\aks$ is the annihilation operator of photons with wave vector $\k$, frequency $\omega_k$, and polarization mode $s$. On the one hand, the substitution of $\ket{\psi(t)}$ on the left-hand side of Eq.~(\ref{Eq:Schrodinger_eq}) yields
\begin{equation} \label{EqSM:susbtituting_ansatz_left}
\begin{aligned}
\begin{split}
    i \frac{d}{dt}\ket{\psi (t)} &= i\frac{d}{dt}(c^{ee}) \ket{ee}\vac +i \sum_{\k,s} \frac{d}{dt}(c_{\ks}^{eg}) \ket{eg} \adks \vac \\
    &+i \sum_{\k,s}\frac{d}{dt}(c_{\ks}^{ge}) \ket{ge} \adks \vac \\
    &+i \sum_{\k,s} \sum_{\substack{(\k', s')\\ \geq(\k, s)}}\frac{d}{dt}(c_{\ks , \ksp}^{gg}) \ket{gg} \adks \adkpsp \vac ,
\end{split}
\end{aligned}
\end{equation}
where all the probability amplitudes are evaluated at time $t$. On the other hand, substituting $\ket{\psi(t)}$ on the right-hand side of Eq.~(\ref{Eq:Schrodinger_eq}) we find
\begin{widetext}
\begin{equation} \label{EqSM:susbtituting_ansatz_right}
\begin{aligned}
    \begin{split}
       \frac{1}{\hbar} H_{I}(t) \ket{\psi(t)} &= \sum_{\k,s} e^{i(\omega_0 - \omega_{k})t}  \biggr[ \dip_{2} \cdot \gks^{(2)} c^{eg}_{\ks } + \dip_{1} \cdot \gks^{(1)} c^{ge}_{\ks } \biggr] \ket{ee}  \vac  \\
        &+\sum_{\k'',s''}     e^{-i(\omega_0 - \omega_{k''})t} \biggr[(\dip_{1} \cdot \gkspp^{(1)})^* \ket{ge}  + (\dip_{2} \cdot \gkspp^{(2)})^* \ket{eg} \biggr] c^{ee} \adkspp \vac  \\
        &+ \sum_{\k,s} \sum_{\substack{(\k', s')\\ \geq(\k, s)}} \sum_{\k'',s''}  e^{i(\omega_0 - \omega_{k''})t} \biggr[\dip_{1} \cdot \gks^{(1)} \ket{eg} + \dip_{2} \cdot \gks^{(2)} \ket{ge} \biggr] {c}^{gg}_{\ks ,\ksp} \akspp  \adks \adkpsp \vac \\
        &+ \sum_{\k,s} \sum_{\substack{(\k', s')\\ \geq(\k, s)}} \sum_{\k'',s''}  e^{-i(\omega_0 - \omega_{k''})t}  \biggr[ (\dip_{1} \cdot \gks^{(1)})^*  c^{eg}_{\ks } + (\dip_{2} \cdot \gks^{(2)})^*  c^{ge}_{\ks }\biggr] \ket{gg} \adkspp \adks  \vac .
    \end{split}
\end{aligned}
\end{equation}
Matching Eqs. (\ref{EqSM:susbtituting_ansatz_left}) and (\ref{EqSM:susbtituting_ansatz_right}) and taking into account that $[\adks, \aksp]=\delta_{\k,\k'}\delta_{s,s'}$ and $[\adks, \adkpsp]=0$, we obtain the set of coupled differential equations 
\begin{subequations} \label{Eq:coupled_differential_eqs}
\allowdisplaybreaks
\begin{align}
    i\frac{d}{dt}c^{ee}(t) &= \sum_{\k, s} c^{ge}_{\k s}(t) \dip_1\cdot \gks^{(1)} e^{i(\omega_0 - \omega_{k})t}+ \sum_{\k , s} c^{eg}_{\k s}(t) \dip_2\cdot \gks^{(2)} e^{i(\omega_0 - \omega_k)t} , \label{eq:dynamics_c_ee} \\
    i\frac{d}{dt}c^{eg}_{\k s}(t) &= e^{-i(\omega_0 - \omega_k)t}c^{ee}(t)(\dip_{2} \cdot\gks^{(2)})^* + \sum_{\k', s'} e^{i(\omega_0 - \omega_{k'})t}(\dip_1\cdot\gkpsp^{(1)}) c^{gg}_{\ks, \ksp}(t) \epsilon (\ks , \ksp) , \label{eq:dynamics_c_eg}  \\
    i\frac{d}{dt}c^{ge}_{\ks}(t) &= e^{-i(\omega_0 - \omega_{k})t}c^{ee}(t)(\dip_{1} \cdot \gks^{(1)})^* + \sum_{\k' ,s'} e^{i(\omega_0 - \omega_{k'})t}(\dip_2 \cdot\gkpsp^{(2)}) c^{gg}_{\ks, \ksp}(t) \epsilon (\ks , \ksp) , \label{eq:dynamics_c_ge}  \\
     i\frac{d}{dt}c^{gg}_{\ks, \ksp}(t) &= \frac{1}{\epsilon (\ks , \ksp)} \biggr[c^{eg}_{\ks}(t) (\dip_{1} \cdot \gkpsp^{(1)})^*e^{-i(\omega_0 - \omega_{k'})t} + c^{eg}_{\ksp}(t) (\dip_{1} \cdot \gks^{(1)})^*e^{-i(\omega_0 - \omega_{k})t} \biggr]  \nonumber\\
     &+ \frac{1}{\epsilon (\ks , \ksp)} \biggr[c^{ge}_{\ks}(t) (\dip_{2} \cdot \gkpsp^{(2)})^*e^{-i(\omega_0 - \omega_{k'})t} + c^{ge}_{\ksp}(t) (\dip_{2} \cdot \gks^{(2)})^*e^{-i(\omega_0 - \omega_{k})t}\biggr]    \label{eq:dynamics_c_gg}.
\end{align}
\end{subequations}
\end{widetext}
We have checked that reducing the above system of differential equations to the case of a single polarization mode $s$ leads to the same system of equations provided in Ref. \cite{Ernst_PR_1968}. 

We use the Wigner-Weisskopf approximation to solve the set of differential coupled  equations in Eq. (\ref{Eq:coupled_differential_eqs}). The first step of this approximation consists in formally integrating the differential equation of the two-photon probability amplitude $c_{\ks,\ksp}^{gg}(t)$ given in Eq.~(\ref{eq:dynamics_c_gg}), which yields
\begin{equation} \label{EqSM:c_gg_formally_integrated}
\begin{split}
    c^{gg}_{\ks, \ksp} (t) &= -\frac{i}{\epsilon (\ks , \ksp) }\int_0^{t} dt'  e^{-i(\omega_0 - \omega_{k'})t'}\\
    &\times \biggr[ 
    c^{eg}_{\ks} (t') (\dip_{1} \cdot \gkpsp^{(1)})^* + c^{ge}_{\ks} (t') (\dip_{2} \cdot \gkpsp^{(2)})^*  \biggr] \\
    &-\frac{i}{\epsilon (\ks , \ksp) }\int_0^{t} dt'  e^{-i(\omega_0 - \omega_{k})t'}\\
    &\times \biggr[c^{eg}_{\ksp}(t') (\dip_{1} \cdot \gks^{(1)})^* + c^{ge}_{\ksp} (t') (\dip_{1} \cdot \gks^{(2)})^*  \biggr]. 
\end{split}
\end{equation}
The next step consists in substituting the above equation into the differential equations of $c^{eg}_{\ks}(t)$ and $c^{ge}_{\ks}(t)$, which are given in Eqs. (\ref{eq:dynamics_c_eg}) and  (\ref{eq:dynamics_c_ge}), respectively. Here, we describe in detail the procedure followed after the substitution into the differential equation of  $c^{eg}_{\ks}(t)$ [an identical procedure is followed after the substitution into the differential equation of $c^{ge}_{\ks}(t)$]. The result of this substitution is
\begin{equation} \label{EqSM:substituting_cgg_into_ceg}
    \begin{split}
        i\frac{d}{dt}c^{eg}_{\ks} (t)&= e^{-i(\omega_0 - \omega_{k})t}c^{ee}(t)(\dip_{2} \cdot\gks^{(2)})^* \\
        &-i \sum_{\ksp} e^{i(\omega_0 - \omega_{k'})t}(\dip_1\cdot\gkpsp^{(1)}) \\
        &\times \biggr[\int_0^{t} dt' c^{eg}_{\ks} (t') (\dip_{1} \cdot \gkpsp^{(1)})^* e^{-i(\omega_0 - \omega_{k'})t'} \\
        &+ \int_0^{t} dt'c^{ge}_{\ks} (t') (\dip_{2} \cdot \gkpsp^{(2)})^* e^{-i(\omega_0 - \omega_{k'})t'}   \\
        &+ \int_0^{t} dt' c^{eg}_{\ksp}(t') (\dip_{1} \cdot \gks^{(1)})^* e^{-i(\omega_0 - \omega_{k})t'} \\
        &+ \int_0^{t} dt' c^{ge}_{\ksp} (t') (\dip_{2} \cdot \gks^{(2)})^* e^{-i(\omega_0 - \omega_{k})t'} \biggr] .
    \end{split}
\end{equation}
The terms in the last two lines of the above expression vanish, as demonstrated in Ref. \cite{Raymond_LP_2007}. After this demonstration, the authors in Ref. \cite{Raymond_LP_2007} consider the simplified case in which both transition dipole moments have identical polarization, which is not our case. 

Next, we take two usual assumptions in the Wigner-Weisskopf approximation \cite{Novotny_book_2012, Milonni_book, Steck_book_2007}. Namely, we assume that (i) the probability amplitudes vary very slowly in time and (ii) the spectral response of the electromagnetic field is very broad. In this way, the decay of the emitter can be interpreted as a Markovian process and the probability amplitudes $c^{eg}_{\ks} (t')$ and $c^{ge}_{\ks} (t')$ in Eq. (\ref{EqSM:substituting_cgg_into_ceg}) can be replaced by $c^{eg}_{\ks} (t)$ and $c^{ge}_{\ks} (t)$, which allows us to take them out of the time integral. Additionally, the upper limit of the integral can be extended to $\infty$. As a consequence, Eq.~(\ref{EqSM:substituting_cgg_into_ceg}) becomes
\begin{equation} \label{EqSM:ceg_dynamics_after_markovian}
\begin{aligned}
    \begin{split}
        &i\frac{d}{dt}c^{eg}_{\ks} (t) = e^{-i(\omega_0 - \omega_{k})t}c^{ee}(t)(\dip_{2} \cdot\gks^{(2)})^* \\
        &-i \sum_{\ksp}  \int_{0}^{\infty} dt' e^{i(\omega_0 - \omega_{k'})(t-t')} \\
        &\times\biggr[c^{eg}_{\ks} (t) |\dip_{1} \cdot \gkpsp^{(1)}|^2  + c^{ge}_{\ks} (t) (\dip_1\cdot\gkpsp^{(1)})(\dip_{2} \cdot \gkpsp^{(2)})^* \biggr].
    \end{split}
\end{aligned}
\end{equation}
At this point, we transform the summation over $\k$ into an integral in the $\k$-space, according to \cite{Milonni_book}
\begin{equation}
    \sum_{\k , s} \rightarrow \frac{\mathcal{V}}{(2\pi)^3} \sum_{s} \int_0^{2\pi}d\phi \int_{0}^{\pi} d\theta \sin\theta \int_{0}^{\infty} dk k^2 .
\end{equation}
Additionally, the calculation of the time integrals in Eq.~(\ref{EqSM:ceg_dynamics_after_markovian}) is facilitated by the Heitler function
\begin{equation}
    \int_{0}^{\infty} dt' e^{i(\omega_0 - \omega_{k})(t-t')} = \pi\delta(\omega_0 - \omega_{k}) +i\mathcal{P}\frac{1}{(\omega_0 - \omega_{k})} ,
\end{equation}
with $\mathcal{P}$ the principal value of the $k$-integral. In this way, Eq.~(\ref{EqSM:ceg_dynamics_after_markovian}) becomes
\begin{equation} \label{EqSM:ceg_dynamics_after_Heitler}
    \begin{split}
        i\frac{d}{dt}c^{eg}_{\ks} &= e^{-i(\omega_0 - \omega_{k})t}c^{ee}(\dip_{2} \cdot\gks^{(2)})^* \\
        &-i \frac{\mathcal{V}}{(2\pi)^3} \sum_{s'=1,2} \int_0^{2\pi}d\phi' \int_{0}^{\pi} d\theta' \sin\theta'  \int_{0}^{\infty} dk' (k')^2    \\
        &\times \biggr[\pi\delta(\omega_0 - \omega_{k'}) +i\mathcal{P}\frac{1}{(\omega_0 - \omega_{k'})}\biggr]   \\
        &\times \biggr[c^{eg}_{\ks}  |\dip_{1} \cdot \gkpsp^{(1)}|^2 
        + c^{ge}_{\ks}  (\dip_1\cdot\gkpsp^{(1)})(\dip_{2} \cdot \gkpsp^{(2)})^*\biggr],
    \end{split}
\end{equation}
where all the probability amplitudes in the above expression are evaluated at time $t$. Next, we decompose the integral in the $\k$-space on the right-hand side of Eq.~(\ref{EqSM:ceg_dynamics_after_Heitler}) into different contributions, which are calculated separately. Each of these contributions emerges from the multiplication of the different terms in brackets inside the integral in Eq.~(\ref{EqSM:ceg_dynamics_after_Heitler}). First, the term proportional to $|\dip_{1} \cdot \gkpsp^{(1)}|^2 \mathcal{P}\{(\omega_0 - \omega_{k'})^{-1}\}$ is ignored because it provides the Lamb-shift induced by the free-space electromagnetic field in the transition frequency of emitter $j=1$ \cite{Milonni_PRA_1974}. This frequency shift is negligible for emitter transitions at optical frequencies and, additionally, the transition frequencies estimated from experiments (e.g., via a one-photon spectrum) include this shift. Thus, we consider that $\omega_0$ in our model already contains this small shift. Second, the term proportional to $|\dip_{1} \cdot \gkpsp^{(1)}|^2 \delta(\omega_0 - \omega_{k'})$ provides the spontaneous emission rate $\gamma_0$ of emitter $j=1$. To demonstrate this, we use the general polarization vectors \begin{subequations}
\allowdisplaybreaks
    \begin{align}
        \hat{\boldsymbol{e}}_{\k 1} &= -\cos\theta\cos\phi \hat{\boldsymbol{x}} -\cos\theta\sin\phi \hat{\boldsymbol{y}} + \sin\theta \hat{\boldsymbol{z}}, \\
        \hat{\boldsymbol{e}}_{\k 2} &= \sin\phi \hat{\boldsymbol{x}} -\cos\phi \hat{\boldsymbol{y}}.
    \end{align}
\end{subequations} 
In this way, we can write
\pagebreak
\begin{widetext}
\begin{equation} \label{EqSM:integral_leading_to_gamma0}
\begin{split}
    &\frac{\mathcal{V}}{(2\pi)^3} \sum_{s'=1,2} \int_0^{2\pi}d\phi' \int_0^{\pi}d\theta' \sin\theta'  \int_0^{\infty} dk' (k')^2    |\dip_{1} \cdot \gkpsp^{(1)}|^2  \pi \delta(\w_i - \omega_{k'})\\
    &=\frac{\mathcal{V}}{(2\pi)^3} \int_0^{\infty} dk' (k')^2 \frac{\omega_{k'}}{2\varepsilon_0 n^2 \hbar \mathcal{V}} \pi \delta(\w_i - \omega_{k'})  \int_0^{2\pi}d\phi' \int_0^{\pi}d\theta' \sin\theta' \sum_{s'=1,2}   |\dip_i \cdot \hat{\boldsymbol{e}}_{\ksp}|^2  \\
    &=\pi \frac{\mathcal{V}|\dip_i|^2}{(2\pi)^3} \int_0^{\infty} dk' (k')^2 \frac{\wk}{2\varepsilon_0 n^2\hbar \mathcal{V}} \delta(\w_i - \omega_{k'})  \int_0^{2\pi}d\phi \int_0^{\pi}d\theta' \sin\theta' \\
    &\times [\cos^2\alpha_i (\cos^2 \theta' \cos^2 \phi' + \sin^2 \phi') + \sin^2 \alpha_i \sin^2 \theta' - 2 \cos\alpha_i \sin\alpha_i \cos\theta' \sin\theta'\cos\phi' ]  \\
    &=\pi \frac{\mathcal{V}|\dip_i|^2}{(2\pi)^3} \frac{8\pi}{3}  \int_0^{\infty} dk' (k')^2 \frac{\omega_{k'}}{2\varepsilon_0 n^2\hbar \mathcal{V}} \delta(\w_i - \omega_{k'}) = \frac{\omega_0^3 |\dip_i|^2}{6\pi \varepsilon_0 n^2 \hbar c^3} = \frac{\gamma_0}{2}.
\end{split}
\end{equation}
The remaining two contributions of the integral in Eq.~(\ref{EqSM:ceg_dynamics_after_Heitler}) are obtained integrating the term proportional to $(\dip_1\cdot\gkpsp^{(1)})(\dip_{2} \cdot \gkpsp^{(2)})^*$. In this case, we find
\begin{equation} \label{EqSM:integral_leading_to_couplings_before_changing_limits}
\begin{split}
    &\frac{\mathcal{V}}{(2\pi)^3} \sum_{s=1,2} \int_0^{2\pi}d\phi \int_{0}^{\pi} d\theta \sin\theta \int_{0}^{\infty} dk k^2  (\dip_1\cdot\gks^{(1)})(\dip_{2} \cdot \gks^{(2)})^* [\pi \delta (\w_0 - \w_k) + i\mathcal{P}\frac{1}{\w_0 - \w_k}]   \\
     &= \frac{\mathcal{V}}{(2\pi)^3} \frac{c  |\dip|^2}{2\varepsilon_0 n^2 \hbar \mathcal{V}} 4\pi \int_{0}^{\infty} dk k^3  [\pi \delta (\w_0 - \w_k) + i\mathcal{P}\frac{1}{\w_0 - \w_k}]
      \\
     &\times  \biggr[\cos\alpha_1 \cos\alpha_2 \biggr(-\frac{\sin(k r_{12})}{(k r_{12})^3} + \frac{\cos(k r_{12})}{(k r_{12})^2} + \frac{\sin(k r_{12})}{(k r_{12})}\biggr) + 2\sin\alpha_1 \sin\alpha_2 \biggr(\frac{\sin(k r_{12})}{(k r_{12})^3} - \frac{\cos(k r_{12})}{(k r_{12})^2} \biggr)  \biggr]   \\
     &= \frac{3 \gamma_0}{4\pi\omega_0^3}\int_{0}^{\infty} d\omega_k \omega_k^3  [\pi \delta (\w_0 - \w_k) + i\mathcal{P}\frac{1}{\w_0 - \w_k}]
      \\
     &\times  \biggr[\cos\alpha_1 \cos\alpha_2 \biggr(-\frac{\sin(k r_{12})}{(k r_{12})^3} + \frac{\cos(k r_{12})}{(k r_{12})^2} + \frac{\sin(k r_{12})}{(k r_{12})}\biggr) + 2\sin\alpha_1 \sin\alpha_2 \biggr(\frac{\sin(k r_{12})}{(k r_{12})^3} - \frac{\cos(k r_{12})}{(k r_{12})^2} \biggr)  \biggr].
\end{split}
\end{equation}
\end{widetext}
The direct calculation of the frequency integral in the above expression does not give the exact dipole-dipole coupling $V$. This discrepancy is due to  the terms ignored by the RWA in the interaction Hamiltonian (which are proportional to $\s_j \aks$ and to $\sd_j \adks$). The RWA does not affect the spontaneous emission rate $\gamma_0$ and the dissipative coupling $\gamma_{12}$ induced by the interaction of the emitters with the free-space electromagnetic field, but it does modify the dipole-dipole coupling $V$ \cite{Agarwal_book_2006, Milonni_PRA_1974}. (This can be checked, for example, by applying the Markovian approximation to trace out the free-space electromagnetic field and obtaining a Markovian master equation in the reduced Hilbert space of the emitters \cite{Stokes_NJP_2018}.) In principle, one could avoid this issue considering the complete interaction Hamiltonian, but the application of the WWA becomes not practical in such case because a generalized ansatz would be required (including, for example, terms with two photons in the field and the two emitters in the excited state). Such a generalized ansatz does not yield a closed system of coupled differential equations for the probability amplitudes. Fortunately, a simple solution to this problem was pointed out by Milonni and Knight in Ref. \cite{Milonni_PRA_1974}: the RWA and the WWA can be safely applied (as we do here) by extending the lower limit of the $k$-integral from $0$ to $-\infty$, which yields the rigorous expression of the coherent dipole-dipole coupling $V$. Following this argument in Eq.~(\ref{EqSM:integral_leading_to_couplings_before_changing_limits}), we obtain
\begin{widetext}
\begin{equation} \label{EqSM:integral_leading_to_couplings}
\begin{split}
     &\frac{3 \gamma_0}{4\pi\omega_0^3}\int_{-\infty}^{\infty} d\omega_k \omega_k^3  [\pi \delta (\w_0 - \w_k) + i\mathcal{P}\frac{1}{\w_0 - \w_k}]
      \\
     &\times  \biggr[\cos\alpha_1 \cos\alpha_2 \biggr(-\frac{\sin(k r_{12})}{(k r_{12})^3} + \frac{\cos(k r_{12})}{(k r_{12})^2} + \frac{\sin(k r_{12})}{(k r_{12})}\biggr) + 2\sin\alpha_1 \sin\alpha_2 \biggr(\frac{\sin(k r_{12})}{(k r_{12})^3} - \frac{\cos(k r_{12})}{(k r_{12})^2} \biggr)  \biggr] \\
     &= \frac{3 \gamma_0}{4} i \biggr[ -\cos\alpha_1 \cos\alpha_2 \frac{\cos(k_0 r_{12})}{(k_0 r_{12})} + (\cos\alpha_1 \cos\alpha_2 -2\sin\alpha_1 \sin\alpha_2) \biggr(\frac{\sin(k_0 r_{12})}{(k_0 r_{12})^2} + \frac{\cos(k_0 r_{12})}{(k_0 r_{12})^3 }\biggr) \biggr] \\
     &+\frac{3 \gamma_0}{4} \biggr[ \cos\alpha_1 \cos\alpha_2 \frac{\sin(k_0 r_{12})}{(k_0 r_{12})} + (\cos\alpha_1 \cos\alpha_2 -2\sin\alpha_1 \sin\alpha_2) \biggr(\frac{\cos(k_0 r_{12})}{(k_0 r_{12})^2} - \frac{\sin(k_0 r_{12})}{(k_0 r_{12})^3 }\biggr) \biggr] \\
     &= iV + \frac{\gamma_{12}}{2} ,
\end{split}
\end{equation}
\end{widetext}
where in the evaluation of the complex integral we have used \cite{Milonni_PRA_1974}
\begin{equation}
\begin{split}
    &\int_{-\infty}^{\infty} d\omega_k \omega_k^3  [\pi \delta (\w_0 - \w_k) + i\mathcal{P}\frac{1}{\w_0 - \w_k}] \\
    &\times\biggr(q\frac{\sin(k r_{12})}{(k r_{12})^3} - q\frac{\cos(k r_{12})}{(k r_{12})^2} +p \frac{\sin(k r_{12})}{(k r_{12})}\biggr) \\
    &= \pi\omega_0^3\biggr(q\frac{\sin(k_0 r_{12})}{(k_0 r_{12})^3} - q\frac{\cos(k_0 r_{12})}{(k_0 r_{12})^2} +p \frac{\sin(k_0 r_{12})}{(k_0 r_{12})}\biggr) \\
    &-i\pi\omega_0^3\biggr(q\frac{\cos(k_0 r_{12})}{(k_0 r_{12})^3} + q\frac{\sin(k_0 r_{12})}{(k_0 r_{12})^2} +p \frac{\cos(k_0 r_{12})}{(k_0 r_{12})}\biggr).
\end{split}
\end{equation}
Additionally, in Eq.~(\ref{EqSM:integral_leading_to_couplings}) we have identified the coherent dipole-dipole coupling \cite{Stokes_NJP_2018}
\begin{equation}
\begin{split}
    V&=\frac{3 \gamma_0}{4} \biggr[ -\cos\alpha_1 \cos\alpha_2 \frac{\cos(k_0 r_{12})}{(k_0 r_{12})} \\
    &+ (\cos\alpha_1 \cos\alpha_2 -2\sin\alpha_1 \sin\alpha_2) \\
    &\times\biggr(\frac{\sin(k_0 r_{12})}{(k_0 r_{12})^2} + \frac{\cos(k_0 r_{12})}{(k_0 r_{12})^3 }\biggr) \biggr]
\end{split}
\end{equation}
and the dissipative coupling
\begin{equation}
    \begin{split}
        \gamma_{12}&=\frac{3 \gamma_0}{2} \biggr[ \cos\alpha_1 \cos\alpha_2 \frac{\sin(k_0 r_{12})}{(k_0 r_{12})} \\
        &+ (\cos\alpha_1 \cos\alpha_2 -2\sin\alpha_1 \sin\alpha_2) \\
        &\times \biggr(\frac{\cos(k_0 r_{12})}{(k_0 r_{12})^2} - \frac{\sin(k_0 r_{12})}{(k_0 r_{12})^3 }\biggr) \biggr] .
    \end{split}
\end{equation}
Substituting Eqs. (\ref{EqSM:integral_leading_to_gamma0}), (\ref{EqSM:integral_leading_to_couplings_before_changing_limits}) and (\ref{EqSM:integral_leading_to_couplings}) into (\ref{EqSM:ceg_dynamics_after_Heitler}) we obtain
\begin{equation} \label{EqSM:c_eg_still_depending_on_c_ee}
    \begin{split}
        i\frac{d}{dt}c^{eg}_{\ks} (t) &= e^{-i(\omega_0 - \omega_{k})t}c^{ee}(t)(\dip_{2} \cdot\gks^{(2)})^* \\
        &-i c^{eg}_{\ks} (t) \frac{\gamma_0}{2}  + c^{ge}_{\ks} (t) (V -i\frac{\gamma_{12}}{2}).
    \end{split}
\end{equation}
Similarly, after applying the same procedure to the differential equation of $c^{ge}_{\ks}(t)$ in Eq.~(\ref{eq:dynamics_c_ge}), we find 
\begin{equation} \label{EqSM:c_ge_still_depending_on_c_ee}
    \begin{split}
        i\frac{d}{dt}c^{ge}_{\ks} (t) &= e^{-i(\omega_0 - \omega_{k})t}c^{ee}(t)(\dip_{1} \cdot\gks^{(1)})^* \\
        &-i c^{ge}_{\ks} (t) \frac{\gamma_0}{2} + c^{eg}_{\ks} (t) (V -i\frac{\gamma_{12}}{2}) .
    \end{split}
\end{equation}
Therefore, Eqs. (\ref{EqSM:c_eg_still_depending_on_c_ee}) and (\ref{EqSM:c_ge_still_depending_on_c_ee}) provide a pair of differential equations for the probability amplitudes $c^{eg}_{\ks} (t)$ and $c^{ge}_{\ks} (t)$, respectively, that depend only on these same probability amplitudes and on $c^{ee}(t)$. Additionally, the coherent dipole-dipole coupling $V$, the dissipative coupling $\gamma_{12}$, and the spontaneous emission rate $\gamma_0$ appear explicitly in these differential equations. 

Furthermore, we consider that the population of the doubly excited state $\ket{ee}$ decays with rate $2\gamma_0$, which can be verified for example using the Markovian master equation to trace the electromagnetic degrees of freedom and reduce to the Hilbert space of the emitters \cite{Stokes_NJP_2018}. Thus,
\begin{equation} \label{Eq:c_ee_decay_assumption}
    c^{ee}(t)=e^{-\gamma_0 t} c^{ee}(0) = e^{-\gamma_0 t} ,
\end{equation}
which yields $| c^{ee}(t)|^2 = \exp (-2\gamma_0 t)$. Consequently, we obtain the set of coupled differential equations
\begin{subequations}
\allowdisplaybreaks
    \begin{align}
        \frac{d}{dt}c^{eg}_{\ks} (t) &= -ie^{-i(\omega_0 - \omega_{k})t}e^{-\gamma_0 t}(\dip_{2} \cdot\gks^{(2)})^* - c^{eg}_{\ks} (t) \frac{\gamma_0}{2} \nonumber \\ 
        &- (iV+\frac{\gamma_{12}}{2}) c^{ge}_{\ks} (t) , \\
        \frac{d}{dt}c^{ge}_{\ks} (t) &= -ie^{-i(\omega_0 - \omega_{k})t}e^{-\gamma_0 t}(\dip_{1} \cdot\gks^{(1)})^* - c^{ge}_{\ks} (t) \frac{\gamma_0}{2} \nonumber \\
        &- (iV+\frac{\gamma_{12}}{2}) c^{eg}_{\ks} (t).
    \end{align}
\end{subequations}
The above differential equations couple only a pair of probability amplitudes $c^{eg}_{\ks} (t)$ and $c^{ge}_{\ks} (t)$ and can be solved analytically without further approximations [in contrast to the set of infinite coupled differential equations in Eq. (\ref{Eq:coupled_differential_eqs})]. The solution of this system is
\begin{subequations}
    \begin{align} 
        2ic^{eg}_{\ks} (t) &=  S_{\ks}^{(-)} e^{-(\frac{\gamma_0 + \gamma_{12}}{2} + iV)t} - A_{\ks}^{(+)}  e^{-(\frac{\gamma_0 - \gamma_{12}}{2} - iV)t} \nonumber\\
        & -( S_{\ks}^{(-)} -  A_{\ks}^{(+)}) e^{-(\gamma_0 + i(\omega_0 - \omega_{k}))t} , \label{EqSM:c_eg_solution}\\
        2ic^{ge}_{\ks} (t) &=  S_{\ks}^{(-)} e^{-(\frac{\gamma_0 + \gamma_{12}}{2} + iV)t} +  A_{\ks}^{(+)}  e^{-(\frac{\gamma_0 - \gamma_{12}}{2} - iV)t} \nonumber \\
        &-(S_{\ks}^{(-)} + A_{\ks}^{(+)}) e^{-(\gamma_0 - i(\omega_0 - \omega_{k}))t}. \label{EqSM:c_ge_solution}
    \end{align}
\end{subequations}

Finally, we substitute Eqs. (\ref{EqSM:c_eg_solution}) and (\ref{EqSM:c_ge_solution}) into the differential equation of $c^{gg}_{\ks, \ksp}(t)$ [given in Eq.~(\ref{eq:dynamics_c_gg})] and solve the resulting time integral, which yields
\begin{widetext}
\begin{equation}
    \begin{split}
       2 \epsilon(\ks ,\ksp)c_{\ks ,\ksp}^{gg}(t) &=- \biggr(1-e^{-[\frac{\gamma_0 + \gamma_{12}}{2} + i(\omega_0 - \omega_{k'}+V)]t}\biggr)  S_{\ks}^{(-)} S_{\ksp}^{(+)} - \biggr(1-e^{-[\frac{\gamma_0 + \gamma_{12}}{2} + i(\omega_0 - \omega_{k}+V)]t}\biggr) S_{\ksp}^{(-)}S_{\ks}^{(+)} \\
        &+\biggr(1-e^{-[\frac{\gamma_0 - \gamma_{12}}{2} + i(\omega_0 - \omega_{k'}-V)]t}\biggr) A_{\ks}^{(+)} A_{\ksp}^{(-)}+ \biggr(1-e^{-[\frac{\gamma_0 - \gamma_{12}}{2} + i(\omega_0 - \omega_{k}-V)]t}\biggr) A_{\ksp}^{(+)}A_{\ks}^{(-)}  \\
        &+ \biggr(1-e^{-[\gamma_0 + i(2\omega_0 - \omega_{k}-\omega_{k'})]t} \biggr) \biggr[S_{\ks}^{(-)} S_{\ksp k}^{(0)} +
           S_{\ksp}^{(-)} S_{\ks k'}^{(0)} -A_{\ks}^{(+)}A_{\ksp k}^{(0)} - A_{\ksp}^{(+)}A_{\ks k'}^{(0)}\biggr].
    \end{split}
\end{equation}
\end{widetext}
The coefficients $S_{\boldsymbol{k}s}^{(\pm)}$, $A_{\boldsymbol{k}s}^{(\pm)}$, $S_{\boldsymbol{k}sk'}^{(\pm)}$, and $A_{\boldsymbol{k}sk'}^{(\pm)}$ are defined in Eqs. (\ref{Eq:9}), (\ref{Eq:10}), (\ref{Eq:14}) and (\ref{Eq:15}) in the main text. The limit $t\rightarrow \infty$ of the above expression leads to Eq.~(\ref{Eq:cgg_steady_state}) in the main text.

\section{Probability density at $s\neq s'$} \label{Appendix:P_xz}
In Sec. \ref{Section:Entanglement_generation} of the main text we have analyzed the probability density $P (\k , s; \k' , s')$ at directions of emission $\hat{\boldsymbol{k}}=-\hat{\boldsymbol{k}}'=\hat{\boldsymbol{y}}$, focusing on the cases in which both photons have the same polarization $s=s'=\hat{\boldsymbol{x}}$ (see Fig. \ref{Figure:3}a in the main text) or $s=s'=\hat{\boldsymbol{z}}$ (see Fig. \ref{Figure:3}b in the main text). Here, we analyze the probability density $P (\k , s; \k' , s')$ for photons with orthogonal polarizations $s=\hat{\boldsymbol{x}}$ and $s'=\hat{\boldsymbol{z}}$ (fixing again 
the directions of emission at $\hat{\boldsymbol{k}}=-\hat{\boldsymbol{k}}'=\hat{\boldsymbol{y}}$). 

\begin{figure}[h] 
	\begin{center}
		\includegraphics[width=0.48\textwidth]{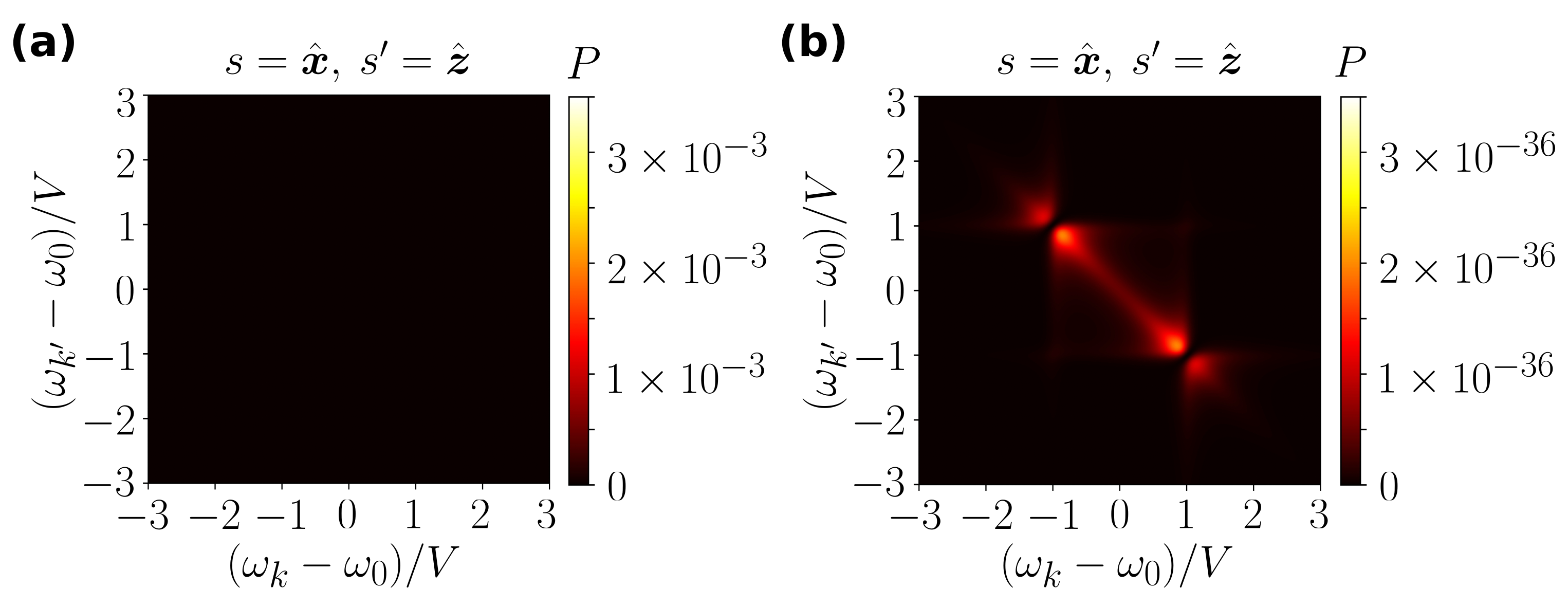}%
		\caption{Analysis of the probability density at $P(\k,s;\k',s')$ for mutually orthogonal polarizations. We plot the dependence of $P(\k,s;\k',s')$ (in units of $\text{m}^2$) at $s=\hat{\boldsymbol{x}}$ and $s'=\hat{\boldsymbol{z}}$, with the directions of emission fixed at $\hat{\boldsymbol{k}}=\hat{\boldsymbol{y}}$ and $\hat{\boldsymbol{k}}'=-\hat{\boldsymbol{y}}$. The emitters considered are two DBATT molecules with perpendicular transition dipole moments, $\gamma_0 /(2\pi)= 21.5$ MHz and $\lambda_0 = 618$ nm. The molecules are assumed to be embedded in a naphthalene crystal with refractive index $n=1.5$ and separated by a distance $r_{12} = 0.075 \lambda_0$.}  
        \label{FigureApp:probaility_density_x_z}  
	\end{center}
\end{figure}

Figure \ref{FigureApp:probaility_density_x_z}a shows the dependence of $P$ on the photon frequencies $\omega_k=kc$ and $\omega_{k'}=k'c$ at $s=\hat{\boldsymbol{x}}$, $s'=\hat{\boldsymbol{z}}$. In this figure, we use a colorbar with the same scale as that in Figs. \ref{Figure:3}a (where $s=s'=\hat{\boldsymbol{x}}$) and \ref{Figure:3}b (where $s=s'=\hat{\boldsymbol{z}}$) in the main text, as well as the same molecular parameters. In this way, we find that the probability density at $s=\hat{\boldsymbol{x}}$, $s'=\hat{\boldsymbol{z}}$ becomes negligible in comparison to the cases discussed in the main text. Further, we modify the scale of the colorbar in Fig. \ref{FigureApp:probaility_density_x_z}b and find that the maximum values of $P$ at $s=\hat{\boldsymbol{x}}$, $s'=\hat{\boldsymbol{z}}$ are $33$ orders of magnitude smaller than the maximum values of $P$ at $s=s'=\hat{\boldsymbol{x}}$ and also at $s=s'=\hat{\boldsymbol{z}}$. Identical behavior of $P$ is obtained at $s=\hat{\boldsymbol{z}}$ and $s'=\hat{\boldsymbol{x}}$. 

Therefore, we conclude that two photons emitted along the directions $\hat{\boldsymbol{y}}$ and $-\hat{\boldsymbol{y}}$ have negligible probability of having mutually orthogonal polarization (in the basis $\hat{\boldsymbol{x}}$ and $\hat{\boldsymbol{z}}$) in comparison to having identical polarizations (in the same basis). 

\section{Combined Debye-Waller/Franck-Condon factor} \label{Appendix:Debye-Waller}
\begin{figure}[t] 
	\begin{center}
		\includegraphics[width=0.45\textwidth]{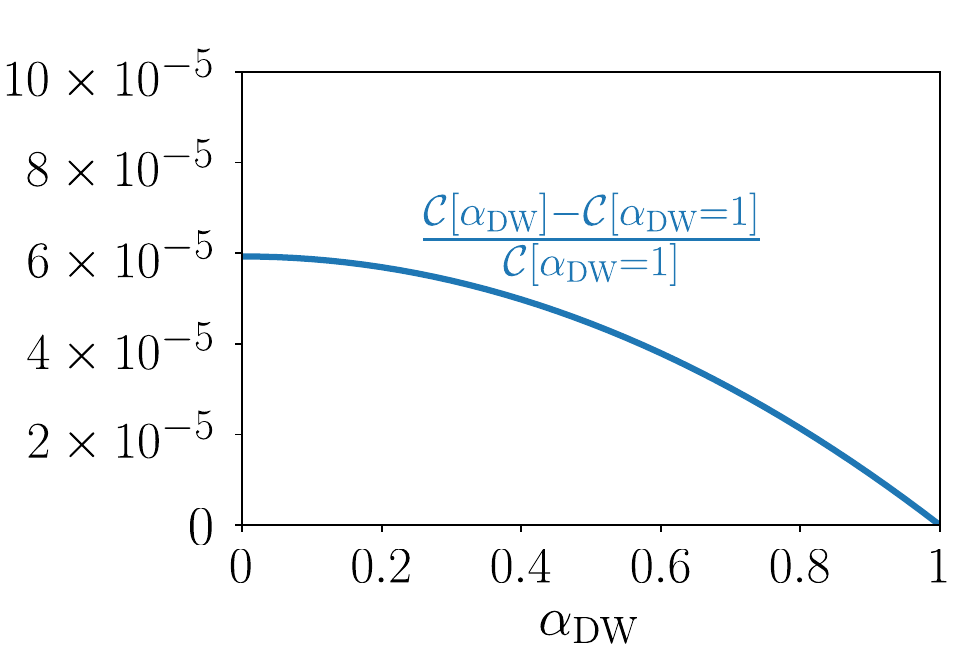}%
		\caption{Influence of the  combined Debye-Waller/Franck-Condon factor $\alpha_{\text{DW}}$ on the entanglement of the two-photon state. We show the dependence of $(\mathcal{C}[\alpha_{\text{DW}}] - \mathcal{C}[\alpha_{\text{DW}}=1])/\mathcal{C}[\alpha_{\text{DW}}=1]$ on $\alpha_{\text{DW}}$, which corresponds to the deviation of the concurrence $\mathcal{C}$ from its value at $\alpha_{\text{DW}}=1$ (which is used in the main text). The coherent dipole-dipole coupling is fixed at $V=11.17 \gamma_0$, which corresponds to $r_{12} = 0.05\lambda_0$ at $\alpha_{\text{DW}}=1$, and the linewidths of the filters at $\Gamma/\gamma_0 = 10^{-2}$. Additionally, we use the parameters $\gamma_0 /(2\pi)= 21.5$ MHz and $\lambda_0 = 618$ nm, which corresponds to DBATT molecules embedded in a naphthalene crystal with refractive index $n=1.5$. }  
        \label{FigureApp:DW}  
	\end{center}
\end{figure}
In this appendix, we show that the combined Debye-Waller/Franck-Condon factor $\alpha_{\text{DW}}$ does not affect significantly the high values of concurrence obtained in Sec. \ref{Section:postselection} of the main text. This $\alpha_{\text{DW}}$ factor is obtained experimentally measuring the ratio of photons emitted in the Zero-Phonon Line from a single emitter (isolated from the interaction with other emitters) over the total number of photons emitted including the Zero-Phonon Line and Stokes-shifted photons \cite{Basche_book}. This factor is thus bounded between $0$ and $1$. 

The theoretical description of the interaction between the two emitters can effectively account for the influence of $\alpha_{\text{DW}}$ by modifying the expressions of the coherent dipole-dipole coupling $V$ and the dissipative coupling $\gamma_{12}$ \cite{Trebbia_NatComms_2022,JuanDelgado}. More specifically, both coupling parameters [which in our case are given in Eqs. (\ref{Eq:coherent_coupling_V}) and (\ref{Eq:dissipative_coupling}) in the main text] are additionally multiplied by $\alpha_{\text{DW}}$. However, as discussed in Sec. \ref{Section:Entanglement_generation} of the main text, the dissipative coupling $\gamma_{12}$ is small in comparison to the spontaneous emission rate $\gamma_0$ for perpendicular transition dipole moments (see the brown line in Fig. \ref{Figure:2}a in the main text, corresponding to the reference configuration in this paper). 
As a consequence, changing $\alpha_{\text{DW}}$ mostly affects the coherent dipole-dipole coupling $V$, in a similar way as changing the distance $r_{12}$ between the emitters. Thus, if we consider $\alpha_{\text{DW}}\neq 1$, the results obtained in the main text can be reproduced to good accuracy by modifying $r_{12}$ appropriately so that $V$ remains fixed according to
\begin{equation} \label{Eq:V_DW}
\begin{split}
    V&=\alpha_{\text{DW}} \frac{3 \gamma_0}{4} \biggr[ -\cos\alpha_1 \cos\alpha_2 \frac{\cos(k_0 r_{12})}{(k_0 r_{12})} \\
    &+ (\cos\alpha_1 \cos\alpha_2 -2\sin\alpha_1 \sin\alpha_2) \\
    &\times\biggr(\frac{\sin(k_0 r_{12})}{(k_0 r_{12})^2} + \frac{\cos(k_0 r_{12})}{(k_0 r_{12})^3 }\biggr) \biggr].
\end{split}
\end{equation}

To verify more rigorously that the influence of $\alpha_{\text{DW}}$ in $\gamma_{12}$ does not alter the high values of concurrence reported in Sec. \ref{Section:postselection} of the main text, we plot in Fig. \ref{FigureApp:DW} the deviation of $\mathcal{C}$ from its value at $\alpha_{\text{DW}}=1$ for different values of combined Debye-Waller/Franck-Condon factor. Additionally, we have fixed $\Gamma=10^{-2}\gamma_0$ and the dipole-dipole coupling at $V=11.17 \gamma_0$ for all $\alpha_{\text{DW}}$, the same as in Fig. \ref{Figure:6} in the main text (where $\alpha_{\text{DW}}=1$ and $r_{12}=0.05\lambda_0$). Thus, $r_{12}$ is changed to maintain this value of $V$ as $\alpha_{\text{DW}}$ is modified. As $V$ is fixed, the variation of $\alpha_{\text{DW}}$ in Fig. \ref{FigureApp:DW} only affects the weak dissipative coupling, according to 
\begin{equation}
\begin{split} \label{Eq:dissipative_coupling_DW}
    \gamma_{12}&=\alpha_{\text{DW}}\frac{3 \gamma_0}{2} \biggr[ \cos\alpha_1 \cos\alpha_2 \frac{\sin(k_0 r_{12})}{(k_0 r_{12})} \\
    &+ (\cos\alpha_1 \cos\alpha_2 -2\sin\alpha_1 \sin\alpha_2)\\
    &\times\biggr(\frac{\cos(k_0 r_{12})}{(k_0 r_{12})^2} - \frac{\sin(k_0 r_{12})}{(k_0 r_{12})^3 }\biggr) \biggr].
\end{split}
\end{equation}
Figure \ref{FigureApp:DW} shows that the maximum deviation of the concurrence with respect to the value that it takes at $\alpha_{\text{DW}}=1$ is $\approx 6\times 10^{-5}$ and occurs at $\alpha_{\text{DW}}\rightarrow0$. Thus, the only notable effect of $\alpha_{\text{DW}}$ for perpendicular transition dipole moments is a change of $V$ that is equivalent to a change of $r_{12}$, as commented in the main text.

\section{Purity of the postselected state} \label{Appendix:Purity}
\begin{figure}[t] 
	\begin{center}
		\includegraphics[width=0.45\textwidth]{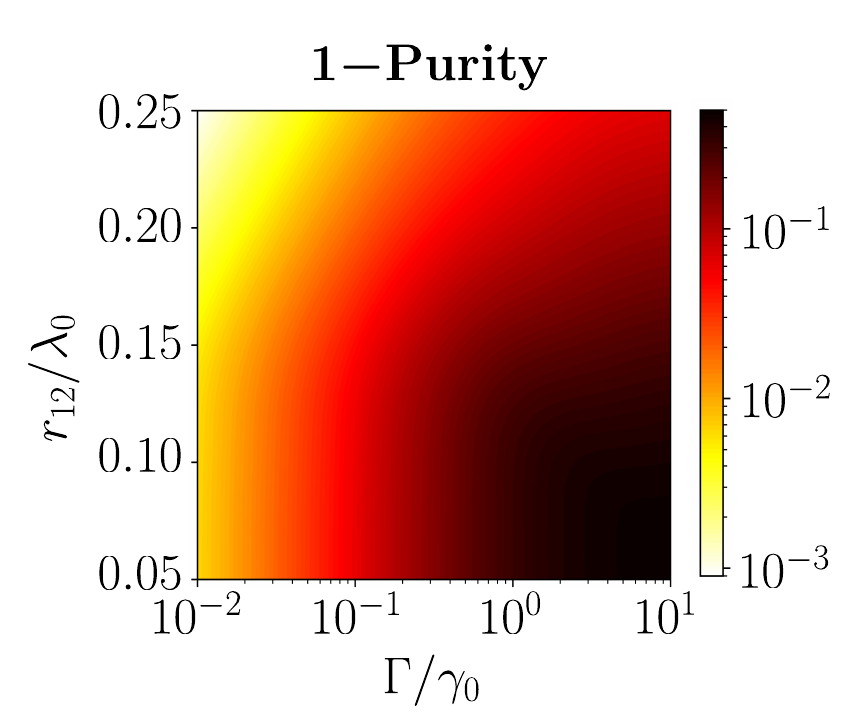}%
		\caption{Analysis of the purity $\mathcal{P}(\hat{\rho})$ of the two-photon postselected state. We plot the dependence of $1-\mathcal{P}(\hat{\rho})$ on the linewidth $\Gamma$ of the filters (normalized by the spontaneous emission rate $\gamma_0$) and on the distance $r_{12}$ between the emitters (normalized by the transition wavelength $\lambda_0$). We consider two DBATT molecules, with $\gamma_0 /(2\pi)= 21.5$ MHz and $\lambda_0 = 618$ nm, embedded in a naphthalene crystal with refractive index $n=1.5$. These two molecules have perpendicular transition dipole moments $\hat{\boldsymbol{\mu}}_1 = (\hat{\boldsymbol{x}}+\hat{\boldsymbol{z}})/\sqrt{2}$ and $\hat{\boldsymbol{\mu}}_2 = (\hat{\boldsymbol{x}}-\hat{\boldsymbol{z}})/\sqrt{2}$.}  
        \label{FigureApp:purity}  
	\end{center}
\end{figure}
In this appendix, we analyze the purity $\mathcal{P}$ of the two-photon postselected state $\hat{\rho}$. As discussed in Sec. \ref{Section:postselection} of the main text, we consider that Alice and Bob use detectors that count all the photons that pass through the optical filters. Thus, the postselected state $\hat{\rho}$ (obtained with the usual tomographic procedure) belongs to the Hilbert space of the polarization of the photons. As the information in the photon frequencies is erased, the two-photon postselected state $\hat{\rho}$ becomes in general a mixed-state. To quantify how much mixed is this postselected state we use $1-\mathcal{P}(\hat{\rho})$, where
\begin{equation}
    \mathcal{P}(\hat{\rho}) = \text{Tr}(\hat{\rho}^2)
\end{equation}
is the standard definition of purity in quantum information. 

We consider the same two DBATT molecules as in the main text and plot in Fig. \ref{FigureApp:purity} the dependence of $1-\mathcal{P}(\hat{\rho})$ on $\Gamma/\gamma_0$ and on $r_{12}/\lambda_0$. We find that the purity of the postselected state decreases with the linewidths of the filters and that $1-\mathcal{P}(\hat{\rho})$ exhibits a similar behavior than that of $1-\mathcal{C}(\hat{\rho})$ in Fig. \ref{Figure:5}a in the main text.

\section{Effect of a lens on the two-photon probability amplitudes} \label{Appendix:detection_angles}
\begin{figure}[t] 
\begin{center}
    \includegraphics[width=0.48\textwidth]{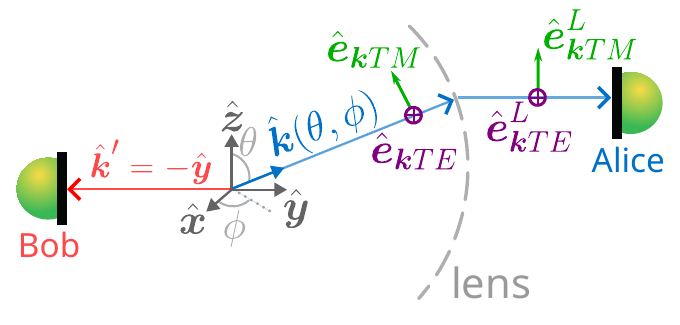}
    \caption{Rotations in the polarization modes of the photon propagating towards Alice produced by a lens oriented perpendicular to the direction $\hat{\boldsymbol{y}}$. The photon detected by Alice is emitted in the direction $\hat{\boldsymbol{k}}=\hat{\boldsymbol{k}}(\theta,\phi)$ (in blue), where $\theta$ and $\phi$ are the polar and azimuthal angles of the wave vector $\k$ (see the Cartesian coordinate system plotted in gray). This photon has orthogonal polarization modes TE (with polarization unit vector 
    perpendicular to the plane of incidence formed by $\hat{\boldsymbol{k}}$ and $\hat{\boldsymbol{y}}$) and TM (with unit vector 
    contained in the plane of incidence). The TE and TM polarization modes have polarization unit vectors $\hat{\boldsymbol{e}}_{\k \text{TE}}$ (in purple) and $\hat{\boldsymbol{e}}_{\k \text{TM}}$ (in green), respectively. The lens (dashed gray line) rotates such unit vectors, so that the photon propagates along the $y$-axis and the polarization becomes contained in the $xz$-plane. The polarization unit vectors after the lens are denoted as $\hat{\boldsymbol{e}}_{\k \text{TE}}^{L}$ and $\hat{\boldsymbol{e}}_{\k \text{TM}}^{L}$. Bob detects a photon propagating in the direction $\hat{\boldsymbol{k}}=-\hat{\boldsymbol{y}}$ (in red).}  
    \label{FigureApp:Lens}  
\end{center}
\end{figure}

In this appendix, we derive the postselected state $\hat{\rho}$ when Alice detects photons emitted at different directions $\hat\k=\hat\k(\theta,\phi)$. As described in Sec. \ref{Subsection:lens} of the main text, we assume that Alice detects light in the forward direction to the $y$-axis [i.e., $\hat\k(\theta,\phi) \cdot \hat{\boldsymbol{y}}>0$] and uses a lens oriented normally to the direction $\hat{\boldsymbol{y}}$, which ensures that light becomes polarized in the $xz$-plane independently of the detection direction $\hat\k(\theta,\phi)$. For simplicity, we consider that the lens is immersed in the same medium (with refractive index $n$) as the quantum emitters. Further, we also assume that Bob detects a photon in the direction $-\hat{\boldsymbol{y}}$ and that both Alice and Bob measure the polarization of photons in the basis formed by the orthogonal directions $\hat{\boldsymbol{x}}$ and $\hat{\boldsymbol{z}}$. We discuss in the following how to model the effect in the two-photon probability  of the lens used by Alice. 

We first recall that the probability amplitude $c_{\ks ,\ksp}^{gg}(\infty)$ [given in Eq.~(\ref{Eq:cgg_steady_state}) in the main text] can be obtained for any two polarization modes (indexed by $s$) with polarization unit vectors $\hat{\boldsymbol{e}}_{\ks}$, as long as these unit vectors are perpendicular to $\k$ and perpendicular between them. [For example, in the calculations performed in Secs. \ref{Section:Entanglement_generation} and \ref{Section:postselection} of the main text we have used the unit vectors $\hat{\boldsymbol{x}}$ and $\hat{\boldsymbol{z}}$ because (i) they are orthogonal to the propagation directions $\hat\k=-\hat\k'=\hat{\boldsymbol{y}}$ considered in these sections, and (ii) Alice and Bob are assumed to measure the polarization in such basis.] Here, in the calculation of the two-photon probability amplitudes we use again the polarization unit vectors $\hat{\boldsymbol{x}}$ and $\hat{\boldsymbol{z}}$ for the photon propagating towards Bob ($\hat\k'=-\hat{\boldsymbol{y}}$), who does not use any lens. In contrast, regarding the photon propagating towards Alice, we choose that its polarization unit vectors are given by the usual transverse-electric (TE) and transverse magnetic (TM) modes (before passing through the lens). Thus, these polarization unit vectors are perpendicular ($\hat{\boldsymbol{e}}_{\k \text{TE}}$) and parallel ($\hat{\boldsymbol{e}}_{\k \text{TM}}$) to the optical plane of incidence, which is formed by the wave vector $\hat\k(\theta,\phi)$ and the direction $\hat{\boldsymbol{y}}$, which is normal to the lens. In this way, these polarization unit vectors before the lens are given by
\begin{subequations}
\begin{align}
    \hat{\boldsymbol{e}}_{\k \text{TE}} &\propto \cos\theta \hat{\boldsymbol{x}} -\sin\theta \cos\phi \hat{\boldsymbol{z}}, \\
    \hat{\boldsymbol{e}}_{\k \text{TM}} &\propto \sin^2 \theta \cos\phi \sin\phi \hat{\boldsymbol{x}} - (\cos^2\theta + \sin^2\theta\cos^2\phi)\hat{\boldsymbol{y}} \nonumber \\
    &+\sin\theta \cos\theta \sin\phi \hat{\boldsymbol{z}} ,
\end{align}
\end{subequations}
as schematically represented in Fig. \ref{FigureApp:Lens}. In the limiting case $\theta\rightarrow\pi/2$ and $\phi\rightarrow\pi/2$, the polarization unit vectors reduce to $\hat{\boldsymbol{e}}_{\k \text{TE}}\rightarrow (\hat{\boldsymbol{x}} - \hat{\boldsymbol{z}})/\sqrt{2}$ and $\hat{\boldsymbol{e}}_{\k \text{TM}}\rightarrow(\hat{\boldsymbol{x}} + \hat{\boldsymbol{z}})/\sqrt{2}$.

This choice of polarization unit vectors of the photon propagating towards Alice facilitates the calculation of the influence of the lens in the two-photon probability amplitudes. On the one hand, the TE-mode is not affected by the lens \cite{Novotny_book_2012}. Thus,
\begin{equation}
    \hat{\boldsymbol{e}}_{\k \text{TE}}^{L}=\hat{\boldsymbol{e}}_{\k \text{TE}} ,
\end{equation}
where the superscript $L$ labels the direction of the polarization after passing through the lens. On the other hand, the TM-mode becomes perpendicular to the $y$-axis and to the TE-mode after passing though the lens \cite{Novotny_book_2012} and can be obtained as
\begin{equation}
    \hat{\boldsymbol{e}}_{\k \text{TM}}^{L} \propto \sin \theta \cos\phi \hat{\boldsymbol{x}} +\cos\theta \hat{\boldsymbol{z}}.
\end{equation}
Importantly, the two-photon probability amplitude $c_{\ks ,\ksp}^{gg}(\infty)$ of the photon propagating towards Alice with polarization $s=\text{TE}$ before passing through the lens coincides with the two-photon probability amplitude after the photon has passed the lens and is polarized in the direction $\hat{\boldsymbol{e}}_{\k \text{TE}}^{L}$. In the same way, the two-photon probability amplitude $c_{\ks ,\ksp}^{gg}(\infty)$ of the photon propagating towards Alice in the TM-mode is equal to the two-photon probability amplitude of the photon polarized in the direction $\hat{\boldsymbol{e}}_{\k \text{TM}}^{L}$ after the lens.

Furthermore, as stated previously, Alice (and also Bob) measures the photon polarization in the basis $\{\ket{\hat{\boldsymbol{x}}},\ket{\hat{\boldsymbol{z}}}\}$. To obtain the two-photon probability amplitudes in such basis a unitary transformation is performed, which can be written as
\begin{equation} \label{Eq:transformation_amplitudes_lens}
\begin{bmatrix}
    c_{\k s=\hat{\boldsymbol{x}} ,\ksp}^{gg}\\
    c_{\k s=\hat{\boldsymbol{z}} ,\ksp}^{gg}
\end{bmatrix}
    = R 
\begin{bmatrix}
    c_{\k s=\text{TE} ,\ksp}^{gg}\\
    c_{\k s=\text{TM} ,\ksp}^{gg} 
\end{bmatrix}.
\end{equation} 
The probability amplitudes on the right-hand side of the above equation can be calculated directly from the analytical expression in Eq.~(\ref{Eq:cgg_steady_state}) in the main text. Importantly, the unitary matrix $R$ is exactly the same one that transforms the polarization unit vectors
\begin{equation}
\begin{bmatrix}
    \hat{\boldsymbol{x}}\\
    \hat{\boldsymbol{z}}
\end{bmatrix}
    = R 
\begin{bmatrix}
    \hat{\boldsymbol{e}}_{\k \text{TE}}^{L}\\
    \hat{\boldsymbol{e}}_{\k \text{TM}}^{L} 
\end{bmatrix}.
\end{equation}
So that, this matrix is given by
\begin{equation}
    R = \frac{1}{\sqrt{\cos^2 \theta + \sin^2 \theta \cos^2 \phi}}
    \begin{bmatrix}
    \cos\theta & \sin\theta \cos\phi\\
    -\sin\theta \cos\phi & \cos\theta
\end{bmatrix}.
\end{equation}
Therefore, the postselected state can be obtained using Eq.~(\ref{Eq:postselected_state}) in the main text and the two-photon probability amplitudes $c_{\k s=\hat{\boldsymbol{x}} ,\ksp}^{gg}$ and $c_{\k s=\hat{\boldsymbol{z}},\ksp}^{gg}$ given by Eq. (\ref{Eq:cgg_steady_state}) in the main text and Eq.~(\ref{Eq:transformation_amplitudes_lens}). 

\section{Estimation of the brightness}\label{Appendix:brightness}
We provide in this appendix a zeroth-order estimation of the brightness associated with the emission of entangled photons from the two interacting emitters. We define this brightness as the number of pairs of photons that can be collected per second and that yield a two-photon postselected state with a high degree of entanglement. To this end, we consider that the quantum emitters are excited using pulsed illumination, with a repetition rate set to one tenth of the spontaneous emission rate $\gamma_0$. For the DBATT molecules discussed in the main text, this corresponds to a total emission rate of $\gamma_0/10 \approx 13.5 \times 10^6$ pairs of photons per second.

We next compute the fraction of these pairs of photons that can be used to postselect a highly entangled two-photon state, which depends on the detection solid angle. Particularly, we focus on the same molecular configuration as in Fig. \ref{Figure:6}a of the main text, where the emitters are separated by $r_{12}=0.05\lambda_0$, their transition dipole moments are mutually orthogonal, and the bandwidth of the filters placed before the detectors is set to $\Gamma=0.01\gamma_0$. From the results in Fig. \ref{Figure:6}a, which shows the fidelity $\mathcal{F}$ with respect to the Bell state $(\ket{\hat{\boldsymbol{x}}\hat{\boldsymbol{x}}}-\ket{\hat{\boldsymbol{z}}\hat{\boldsymbol{z}}})/\sqrt{2}$ as a function of the detection direction of Alice, we estimate that $\mathcal{F}\geq0.99$ within the solid angle $\Omega_{\mathcal{F}\geq0.99}$, approximately defined by the ellipse
\begin{equation}
1 = \left(\frac{\theta - \pi/2}{\pi/5}\right)^2 + \left(\frac{\phi - \pi/2}{\pi/4}\right)^2 .
\end{equation}
We thus consider that only photons emitted within this solid angle are collected. To compute the collection probability within such a solid angle, we first take into account that the power density radiated by a dipole aligned along the $z$-axis [corresponding to the cascade emission through the antisymmetric state $\ket{A}=(\ket{ge}-\ket{eg})/\sqrt{2}$, marked with green arrows in Fig. \ref{Figure:2}b] is given by $\frac{3}{8\pi}\sin^2\theta$ \cite{Balanis_book_2016}. This expression for the power density is normalized by the total radiated power. Thus, by integrating this power density over the region $\Omega_{\mathcal{F}\geq0.99}$ (where $\mathcal{F}\geq0.99$), we obtain that the collection probability is approximately
\begin{equation}
I_{\mathcal{F}\geq0.99} \equiv \int_{\Omega_{\mathcal{F}\geq0.99}} d\Omega \frac{3}{8\pi}\sin^2\theta \approx 0.13 .
\end{equation}
We then consider that the radiation pattern of the second photon and, thus, the collection probability is the same as that of the first one (as it also corresponds to a dipole oriented along the $z$-axis). We thus estimate that $I_{\mathcal{F}\geq0.99}^2 \gamma_0 /10 \approx 2.3 \times 10^5$ entangled photon pairs per second can be collected over such a solid angle. Equivalent results are obtained if both photons are considered to be emitted from the radiation of the symmetric state $\ket{S}=(\ket{ge}+\ket{eg})/\sqrt{2}$.

Finally, we account for the effect of the filtering process on the collection probability. From Fig. \ref{Figure:5}c, we estimate that the probability of collecting an entangled photon pair is reduced by $N/N_{\text{max}}(r_{12}=0.05\lambda_0)\approx 1.25\times10^{-5}$ [with $N_{\text{max}}(r_{12}=0.05\lambda_0)$ the maximum value of $N$ in Fig. \ref{Figure:5}c for $r_{12}=0.05\lambda_0$], as compared with the case of spectrally very broad filters (which is similar to the case of no filters), for the same molecular configuration (with $r_{12}=0.05\lambda_0$) and the same optical filters (with $\Gamma=0.01\gamma_0$). In this way, we obtain that $\frac{\gamma_0}{10} I_{\mathcal{F}\geq0.99}^2 N/N_{\text{max}} \approx 3$ highly entangled photon pairs per second can be obtained after the collection and filtering processes. This value could be increased by relaxing the fidelity threshold ($\mathcal{F}\geq0.99$) and/or by increasing the filter bandwidth, as well as through the use of optical cavities.

\section{Distant emitters}\label{Appendix:distant_emitters}
\begin{figure*}[t] 
\begin{center}
    \includegraphics[width=0.95\textwidth]{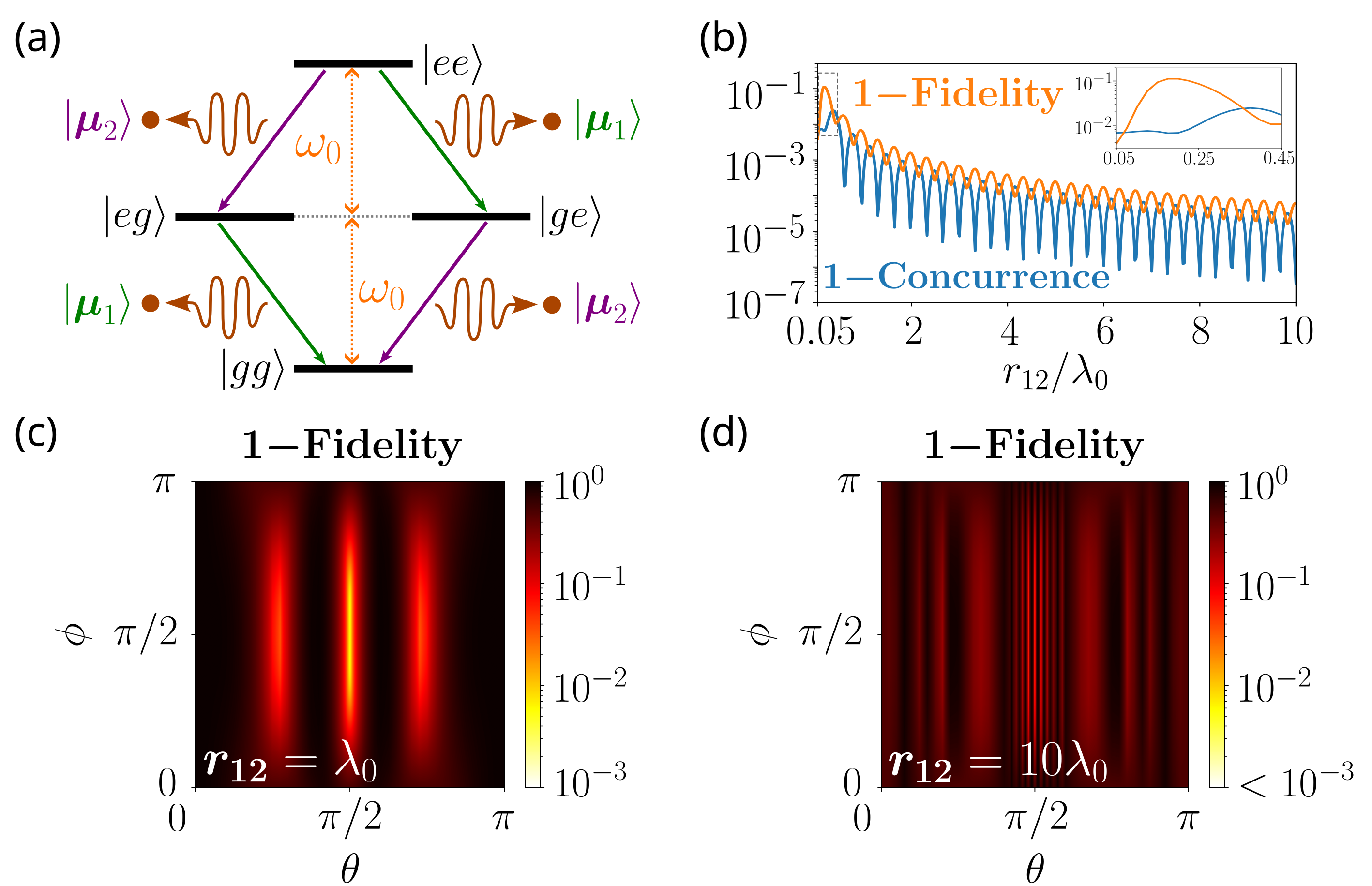}
    \caption{Characterization of the postselected state for increasing distances $r_{12}$ between the emitters. (a) Schematic representation of the energy levels and decay paths for two uncoupled emitters. The emitters are assumed to have identical transition frequencies $\omega_0$ and thus the relaxation of any of the emitters produces a photon at this frequency (in brown). The transition dipole moments of the emitters are oriented perpendicularly to each other, with $\hat{\boldsymbol{\mu}}_1 = (\hat{\boldsymbol{x}}+\hat{\boldsymbol{z}})/\sqrt{2}$ and $\hat{\boldsymbol{\mu}}_2 = (\hat{\boldsymbol{x}}-\hat{\boldsymbol{z}})/\sqrt{2}$. The relaxation of emitter $j$ leads to the emission of a photon with polarization state $\ket{\boldsymbol{\mu}_j}$. (b) Dependence on $r_{12}/\lambda_0$ of $1-\mathcal{C}(\hat{\rho})$ (blue solid line) and of $1-\mathcal{F}(\hat{\rho})$ (orange solid line). In this panel we consider that Alice detects photons emitted in the direction $\hat{\boldsymbol{y}}$ and Bob does it in the direction $-\hat{\boldsymbol{y}}$. In the inset we make a zoom of the behavior of $1-\mathcal{C}(\hat{\rho})$ and $1-\mathcal{F}(\hat{\rho})$ at the interval $0.05 \lambda_0 \leq r_{12} \leq 0.45 \lambda_0$,corresponding to the dashed gray box.  [(c),(d)] Dependence of $1-\mathcal{F}(\hat{\rho})$ [with $\mathcal{F}(\hat{\rho})$ the fidelity with respect to the Bell state $(\ket{\hat{
    \boldsymbol{x}}}_A \ket{\hat{
    \boldsymbol{x}}}_B - \ket{\hat{
    \boldsymbol{z}}}_A \ket{\hat{
    \boldsymbol{z}}}_B )/\sqrt{2}$] on the direction of detection $\hat\k = \hat\k (\theta, \phi)$ of Alice for separation distances (c) $r_{12}=\lambda_0$, and (d) $r_{12}=10\lambda_0$. $\theta$ and $\phi$ are the polar and azimuthal angles of the wave vector $\hat\k$ in spherical coordinates, see the sketch in Fig. \ref{FigureApp:Lens}. The detection of direction of Bob is fixed at $\hat\k' = -\hat{\boldsymbol{y}}$. The minimum value of the colormap in (d) is saturated to facilitate a better comparison with other colormaps in which the behavior of the fidelity is also analyzed. In (b), (c), and (d), we consider that the emitters are DBATT molecules, with $\gamma_0 /(2\pi)= 21.5$ MHz and $\lambda_0 = 618$ nm, which are embedded in a naphthalene crystal with refractive index $n=1.5$. }  
    \label{FigureApp:DistantEmitters}  
\end{center}
\end{figure*}
In this appendix, we analyze the postselected state at larger values of intermolecular distance $r_{12}$ than those analyzed in the main text. We show that the postselected two-photon state generated from the relaxation of two distant emitters can yield high values of concurrence when Alice and Bob detect light propagating at directions $\hat{\boldsymbol{k}}=\hat{\boldsymbol{y}}$ and $\hat{\boldsymbol{k}}'=-\hat{\boldsymbol{y}}$ (as in Sec. \ref{Section:postselection} of the main text), respectively. However, we discuss below how these states may not be well suited for practical experiments.

We first give a simple argument on how, at large separation distances between two quantum emitters with orthogonal transition dipole moments $\boldsymbol{\mu}_1 = \mu(\hat{\boldsymbol{x}}+\hat{\boldsymbol{z}})/\sqrt{2}$ and $\boldsymbol{\mu}_2 = \mu(\hat{\boldsymbol{x}}-\hat{\boldsymbol{z}})/\sqrt{2}$, a highly entangled two-photon state can be again postselected. At sufficiently large separation distance $r_{12}$, the dipole-dipole coupling becomes negligible (see Fig. \ref{Figure:2}a in the main text). Thus, the eigenstates of the system are simply $\ket{g g}$ (with eigenvalue $0$), $\ket{g e}$ (with eigenvalue $\hbar\omega_0$), $\ket{e g}$ (with eigenvalue $\hbar\omega_0$) and $\ket{e e}$ (with eigenvalue $2\hbar\omega_0$), as schematically represented in Fig. \ref{FigureApp:DistantEmitters}a. The radiative decay from the doubly excited state generates two photons at frequencies $\omega_0$, one of them with polarization $\hat{\boldsymbol{\mu}}_1$ and the other one with polarization $\hat{\boldsymbol{\mu}}_2$, as they are generated from the independent relaxation of each emitter. Thus, the postselected state becomes a superposition of two detection possibilities: (i) the photon propagating towards Alice (in the direction $\hat{\boldsymbol{y}}$) having polarization $\hat{\boldsymbol{\mu}}_1$ and the photon propagating towards Bob (in the direction -$\hat{\boldsymbol{y}}$) having polarization $\hat{\boldsymbol{\mu}}_2$, and (ii) the opposite situation, in which the photon propagating towards Alice is polarized in the direction $\hat{\boldsymbol{\mu}}_2$ and the photon propagating towards Bob is polarized in the direction $\hat{\boldsymbol{\mu}}_1$. Thus, we expect that the two-photon state is given as 
\begin{equation}
\begin{split}
    \ket{\psi (r_{12}\rightarrow\infty)} &= \frac{\ket{\hat{\boldsymbol{\mu}}_1}_A \ket{\hat{\boldsymbol{\mu}}_2}_B + \ket{\hat{\boldsymbol{\mu}}_2}_A \ket{\hat{\boldsymbol{\mu}}_1}_B}{\sqrt{2}} \\
    &= \frac{\ket{\hat{
    \boldsymbol{x}}}_A \ket{\hat{
    \boldsymbol{x}}}_B - \ket{\hat{
    \boldsymbol{z}}}_A \ket{\hat{
    \boldsymbol{z}}}_B}{\sqrt{2}} .
\end{split}
\end{equation}
This state is equivalent to the state $\ket{\psi_-^{\text{Bell}}}$ expected for very short separation distances $r_{12}$ (see Secs. \ref{Section:Entanglement_generation} and \ref{Section:postselection} of the main text), although the physical mechanism describing the generation of the photon pair is different, as well as the photon frequencies. 

Next, following the procedure described in Sec. \ref{Section:postselection} of the main text and considering again two DBATT molecules as reference emitters, we calculate the concurrence $\mathcal{C}(\hat{\rho})$ of the postselected state (at $\hat{\boldsymbol{k}}=-\hat{\boldsymbol{k}}'=\hat{\boldsymbol{y}}$) for large separation distances, as well as the fidelity $\mathcal{F}(\hat{\rho})$ with respect to the Bell state $\ket{\psi (r_{12}\rightarrow\infty)}=\ket{\psi_-^{\text{Bell}}}$. Figure \ref{FigureApp:DistantEmitters}b shows $1-\mathcal{C}(\hat{\rho})$ (solid blue line) and $1-\mathcal{F}(\hat{\rho})$ (solid orange line). We find that the concurrence and the fidelity can be optimized in two different ways. On the one hand,  when the coherent dipole-dipole interaction between the emitters is significant (equivalently, at short separation distances $r_{12}$), the photon entanglement generally increases (following an oscillatory behavior) for decreasing values of $r_{12}$ (see the inset in Fig. \ref{FigureApp:DistantEmitters}b), as discussed in the main text. On the other hand, if the dipole-dipole interaction is weak (equivalently, at large separation distances $r_{12}$), we find that the photon entanglement increases overall for larger values of $r_{12}$, which is consistent with the simple argument given in the previous paragraph. As a consequence, we observe in Fig. \ref{FigureApp:DistantEmitters}b that $1-\mathcal{C}(\hat{\rho})$ reaches a maximum value (corresponding to lower photon entanglement) at an intermediate regime of separation distances ($r_{12} \sim 0.35 \lambda_0$), where the dipole-dipole interaction is neither very weak nor very strong.

Finally, we discuss why, despite the high values of concurrence obtained at large separation distances $r_{12}$, we do not expect such a configuration to be practical for applications in quantum technologies. With this purpose, we analyze the two-photon postselected state under different detection directions and at large separation distances. As discussed in Sec. \ref{Subsection:lens} of the main text, we expect that the two-photon state measured in experiments including lenses is highly entangled if large values of fidelity $\mathcal{F}(\hat{\rho})$ with respect to a Bell state are obtained at each direction over the solid angle of collection (given by the numerical aperture of the lens). We consider $r_{12}=\lambda_0$ and plot in Fig. \ref{FigureApp:DistantEmitters}c the dependence of $1-\mathcal{F}(\hat{\rho})$ on the detection direction of Alice (with the detection direction of Bob again fixed at $\hat{\boldsymbol{k}}'=-\hat{\boldsymbol{y}}$), where the fidelity $\mathcal{F}(\hat{\rho})$ with respect to the Bell state $\ket{\psi (r_{12}\rightarrow\infty)}$ is obtained following the procedure described in Sec. \ref{Section:postselection} of the main text and in Appendix~\ref{Appendix:detection_angles}. We find that $1-\mathcal{F}(\hat{\rho})$ drastically increases under small deviations from $\theta=\pi/2$. This deviation becomes more extreme for increasing separation distances, as shown in Fig. \ref{FigureApp:DistantEmitters}d, where we have fixed $r_{12}=10\lambda_0$. In particular, by examining the variation of
$1-\mathcal{F}(\hat{\rho})$ as a function of $\theta$ at fixed $\phi = \pi/2$ in Figs. \ref{FigureApp:DistantEmitters}c and
\ref{FigureApp:DistantEmitters}d, we find that the full width at half maximum of the central dip of $1-\mathcal{F}(\hat{\rho})$ is
approximately $10$ times larger at $r_{12} = \lambda_0$ (Fig. \ref{FigureApp:DistantEmitters}c) than at $r_{12} = 10\lambda_0$ (Fig.
\ref{FigureApp:DistantEmitters}d). Therefore, we expect that obtaining a highly entangled two-photon state from two quantum emitters separated by large distances becomes very challenging in practice, in contrast to the case of short separation distances discussed in Sec. \ref{Subsection:lens} of the main text.   

\bibliography{bibliography_entanglement.bib} 
\end{document}